\documentclass[11pt]{article}
\pdfoutput=1 

\usepackage{jheppub} 

\usepackage[T1]{fontenc} 

\usepackage{amsmath,amssymb}

\usepackage{color}

\newcommand{\pd}{\partial}
\newcommand{\rme}{{\rm e}}

\newcommand{\rmi}{{\rm i}}          

\title{
    Performance of Complex Langevin Simulation in 0+1 dim.\
    massive Thirring model at finite density
  }

\author[a]{Hirotsugu Fujii,}
\author[b,c,d]{Syo Kamata,}
\author[a]{and Yoshio Kikukawa}

\affiliation[a]
{Institute of Physics, University of Tokyo, Komaba, Tokyo 153-8092, Japan}
\affiliation[b]
{Department of Physics, Rikkyo University, Tokyo 171-8501, Japan}
\affiliation[c]
{Department of Physics, Keio University, Yokohama, Japan}
\affiliation[d]
{Physics Department, Fudan University, Shanghai 200433, China}

\emailAdd{hfujii@phys.c.u-tokyo.ac.jp}
\emailAdd{skamata@rikkyo.ac.jp}
\emailAdd{kikukawa@hep1.c.u-tokyo.ac.jp}

\abstract{
Statistical sampling with the complex Langevin (CL) equation is applied
to (0+1)-dimensional Thirring model, and its uniform-field variant,
at finite fermion chemical potential $\mu$.
The CL simulation reproduces a crossover behavior which is similar
to but actually deviating from the exact solution
in the transition region, where we confirm that the CL simulation
becomes susceptible to the drift singularities,
{\it i.e.}, zeros of the fermion determinant.
In order to simulate the transition region
with the CL method correctly,
we examine two approaches, a reweighting method
and a model deformation, in both of which
a single thimble with an attractive fixed point
practically covers the integration domain
and the CL sampling avoids the determinant zeros.
  It turns out that these methods
can reproduce the correct crossover
behavior of the original model with
using reference ensembles in the complexified space.
However, 
they need evaluation of the reweighting factor,
which scales with the system size exponentially.
We discuss feasibility of applying these methods
to the Thirring model and to more realistic theories.
}

\begin{document} 

\maketitle

\flushbottom

\section{Introduction}

Simulating a fermion system at finite chemical potentials
often leads to the notorious sign problem with a complex action
in the path-integral formalism.
Study of QCD phase diagram at finite densities is one of the most
challenging subjects of this category.
Recently, among other extensive attempts\cite{deForcrand:2010ys},
two approaches which involve field complexification have been studied
as a promising direction to the solution of the problem.
Application of the Langevin equation to
a complex action problem is a classic idea
\cite{Parisi:1984cs,Klauder:1983zm,Klauder:1983sp},
and has been attracting renewed attention with 
recent progress
\cite{Aarts:2008rr,Aarts:2008wh,Aarts:2009dg,
Aarts:2009uq,Aarts:2010aq,Aarts:2010gr,Aarts:2011ax,Aarts:2011sf,
Aarts:2011zn, Aarts:2012ft,Aarts:2013uza,
Makino:2015ooa}
made in the last decade
including exploratory applications to QCD at finite density
in \cite{Seiler:2012wz,Sexty:2013ica,Aarts:2014bwa,Aarts:2014kja,Fodor:2015doa},
also in \cite{Nagata:2016mmh}
(see \cite{Aarts:2013uxa,Aarts:2014fsa,Seiler:2017wvd} for review).
Generalization of the steepest-descent-path idea
for oscillatory integrals to the multi-dimensional cases
is called now the Lefschetz-thimble approach
\cite{Cristoforetti:2012su,Cristoforetti:2013wha,Fujii:2013sra} 
and has been applied to simple bosonic theories
\cite{Cristoforetti:2013wha,Fujii:2013sra,Cristoforetti:2014gsa,Tanizaki:2014xba,Tanizaki:2014tua,Tanizaki:2015pua,Alexandru:2016san}
as well as fermionic theories
\cite{Mukherjee:2014hsa,Kanazawa:2014qma,DiRenzo:2015foa,
Tanizaki:2015rda,Fujii:2015bua,Fujii:2015vha,
Alexandru:2015xva,Alexandru:2015sua,Alexandru:2016ejd}.
Furthermore several remarks are found in
\cite{Aarts:2013fpa,Aarts:2014nxa,Fukushima:2015qza,Nishimura:2017vav}
in conjunction with the two approaches.

Long-time evolution of the Langevin equation of real action $S(x)$
\begin{align}
  \frac{dx}{d\tau} = - \frac{\partial S}{\partial x} + \eta(\tau)
\end{align}
with a stochastic noise $\eta(\tau)$
is a standard method to generate the
equilibrium ensemble $P(x)\propto \rme^{-S(x)}$
of configurations $x$, which is shown
by the spectral analysis of the
associated Fokker-Planck operator.
Apparently,
the Langevin evolution can be applied to a complex action system
$S \in {\mathbb C}$,
once the configuration space is complexified:
$x \in {\mathbb R}^n \to z=x+\rmi y \in {\mathbb C}^n$,
which results in the complex Langevin (CL) equation.
One can solve CLE in practice by applying adaptive stepsize
in the Langevin time $\tau$
to suppress numerical instability of the evolution trajectories
\cite{Aarts:2009dg}.

It has been recognized, however, that the resultant distribution
$P(x,y)$ in the complexified space $z=x+\rmi y$
converges to an incorrect solution in some cases, including
fermionic systems at finite density
\cite{SanoLattice2011,Pawlowski:2013pje,Pawlowski:2013gag,
  Mollgaard:2013qra,Mollgaard:2014mga,
  Hayata:2015lzj}.
Correctness of the CL method is argued relying on the holomorphic property
of the associated Fokker-Planck equation
with repeated use of integration by parts,
with requiring the distribution $P(x,y)$ localized in imaginary directions
\cite{Aarts:2009uq,Aarts:2011sf,Aarts:2013uza}.
But the fermion determinant gives rise to a meromorphic drift
force, and the condition for correctness of the method is revisited 
in Refs.~\cite{Nishimura:2015pba,Nagata:2015uga,Nagata:2016vkn,Aarts:2017vrv},
where subtleties with the meromorphic drift term is carefully studied.
It is pointed out that the singularities of the drift force give
additional surface-term contribution in the integration by parts,
and also that they prevent us from taking the continuum limit of the
discretized CL equation,
depending of the support of the CL sample distribution.
It is claimed in Ref.~\cite{Nagata:2016vkn} that
CL method gives the correct results if
the distribution $P(x,y)$
is suppressed at large drift magnitude $|K| \equiv |\partial S/\partial z|$
exponentially,
  $P(x,y) \sim \rme^{-\text{const.} |K(z)|}$ for large $|K|$,
or stronger.

In this work we examine statistical simulations
with the CL equation applied to the (0+1) dimensional
Thirring model, and its uniform-field variant,
at finite chemical potential~$\mu$
\cite{Pawlowski:2013pje,Pawlowski:2013gag,Pawlowski:2014ada,Fujii:2015bua}.
This model is analytically solvable on a finite lattice,
showing a smooth crossover with increasing $\mu$.
It has been used as a good benchmark for
testing new methods for the sign problem in a fermionic system
\cite{Pawlowski:2013pje,Pawlowski:2014ada,Fujii:2015bua,Fujii:2015vha,Alexandru:2015xva}.
The CL simulations are performed in (0+1) dimensional Thirring model
and the results are compared with the analytic results
\cite{Pawlowski:2013pje,Fujii:2015bua},
and the consistency condition are examined\cite{Pawlowski:2013pje}.
It has been known that 
CL simulation is successful at small and large $\mu$ outside the crossover,
but converges to an incorrect solution in between.

Here we confirm numerically that for $\mu$ in the crossover region
the CL simulations of the model becomes susceptible to
the singularities of the drift term and
the ensemble distribution is only
power-suppressed in the vicinities of the determinant zeros,
which makes the method unjustifiable there\cite{Nagata:2016vkn}.
From the structure of the drift flow field of the model,
it seems inevitable that the CL sampling hits the vicinity of the
drift singularities (fermion determinant zeros) for $\mu$
in the crossover region.
At small or large $\mu$, where CL method is successful,
the determinant zeros locate outside the support of the CL ensemble.

In view of the thimble (steepest descent path) integration,
the abrupt crossover behavior is resulted from
the contributions of the multi thimbles connected via the
determinant zeros\cite{Tanizaki:2015rda,Fujii:2015bua,Fujii:2015vha}.
The destructive interference between multi-thimble contributions poses
the global sign problem in numerical evaluation.
In contrast, 
at zero or large $\mu$, a single thimble ${\cal J}_{z_0}$
practically covers the integration domain. In this latter case,
the CL sampling $P(x,y)$ around the thimble
does not hit the determinant zeros.

Several attempts are made so far
in order to extend the applicability of CL simulations
to the crossover region.
One is the reweighting method:
Observables in the crossover region
are computed by reweighting with 
the CL ensembles generated at a reference point
within the (multi-) parameter space
where the CL method is valid
\cite{Bloch:2016jwt,Bloch:2017ods,Bloch:2017sfg},
at the cost of the so-called reweighting factor.
Deformation of the models 
is proposed in \cite{Tsutsui:2015tua,Doi:2017gmk}, where 
the fermion determinant is modified so that the determinant
zeros become irrelevant in the CL sampling, and
observables of the original model
are re-expressed exactly
in terms of those of the deformed models.
  Another model-deformation method
is studied in \cite{Ito:2016efb}
with introduction of an auxiliary parameter
into the fermion determinant
 to study the spontaneous symmetry breaking.
The CL method is applied to a fermionic model
within the applicable parameter region,
and then the zero limit of the parameter
is taken by extrapolation.
The gauge cooling method is extended
in \cite{Nagata:2016alq}
for avoiding the artificial singular drift problem
in a random matrix model.

In this work,
we examine the reweighting method\cite{Bloch:2017ods}
and model-deformation method\cite{Tsutsui:2015tua,Doi:2017gmk}.
The latter may be regarded as a kind of reweighting method,
for it needs evaluation of the reweighting factor, too,
as seen in Eq.~(\ref{eq:TDidentity}) below.
In reweighting we use 
the CL ensembles generated with zero or large $\mu$, for which
the ensemble supports have no overlap
with the vicinity of the zeros and also
the corresponding thimble structure becomes simple.
The reweighting method can reproduce the correct crossover behavior
(for a small system size), 
which enforces the correctness of the CL ensembles
generated outside the crossover region.
Then we study 
the efficiency of the reweighting method by changing the system size.
For deformation method, 
we find a subtlety in dealing with the pole of the fermion observables
such as number and scalar densities.
We will comment on this point,
in addition to the efficiency of the method.

This paper is organized as follows:
In the next section we present the definition of the model
and its uniform-field variant,
and perform CL simulations for them.
We then analyze the success and failure of
CL simulations by inspecting the ensemble distributions on the complex plane
in relation to the zeros, critical points, and thimble structure
in some detail, and make histograms of the drift force to test the
criterion for the validity of CL method. We end Sec.~2 with a brief summary.
In Sec.~3 we apply the re-weighting method to simulate the model in
the crossover region using zero- and large-$\mu$ reference ensembles.
Main focus is the efficiency of the re-weighting w.r.t.\
the system size toward larger systems.
In Sec.~4 we examine the deformation method, where
we encounter a difficulty due to the poles in the observable
and/or numerical efficiency in this approach.
Sec.~5 is devoted to Summary. Appendix A present analytic expressions
for the determinant zeros in the deformed model.

\section{Model}
\subsection{Thirring model in (0+1) dimensions}
We deal with the lattice Thirring model 
with a single staggered ferimon in (0+1) dimensions
at finite chemical potential $\mu$.
After unfolding the vector-type four-point interaction,
we can write the partition function $Z$ 
as a path integral over a compact auxiliary filed $A_n$
\cite{Pawlowski:2013pje,Pawlowski:2013gag,Pawlowski:2014ada,Fujii:2015bua}:
\begin{align}
\label{eq:path-integral-original}
Z =& \int_{-\pi}^{\pi} \prod_{n=1}^L  \frac{d A_n }{2\pi} \, 
     \rme^{- \beta  S_b [A] } 
     \, \det D \, [A;\mu]
\end{align}
with the bosonic action $S_b[A]=\sum_{n=1}^{L}  \big(1- \cos A_n \big)$
and the fermion determinant 
\footnote{
  Note that we have omitted an irrelevant factor of $2^{L-1}$ in
$\det D$ for notational simplicity.
}
\begin{align}
\label{eq:detD-A}
 \det D \, [A;\mu] = 
\cosh (L   \mu + \rmi {\scriptstyle \sum_{n=1}^L } A_n) + \cosh L \hat m 
\, ,
\end{align}
where $L$ is the temporal lattice size,
$\beta=1/(2g^2)$ the inverse coupling constant, $\hat m= \sinh^{-1} m$ the
fermion mass term.
We have set the lattice unit $a=1$ and 
all the dimensionful quantities are measured in the lattice unit
hereafter.
We notice here
that $\text{det}D[A;\mu]$ depends on the field configuration
only through the sum $s\equiv \sum_{n=1}^{L}A_n$.
The sign problem arises from the complex determinant $\det D$
at nonzero $\mu$ in this model.

The analytic expression for $Z$ is known as
\begin{align}
  Z= 
  \rme^{-\beta L}
  \left [
    I_1(\beta)^L\cosh L\mu  +  I_0(\beta)^L \cosh L\hat m
  \right ]
\,  
,
\label{eq:Z}
\end{align}
where $I_{0,1}(x)$ are the modified Bessel function of the first kind.
The fermion number density $\langle n \rangle$ and
scalar density $\langle \sigma \rangle$ are obtained,
respectively, as 
\begin{align}
\langle n \rangle  = &  \frac{1}{L}    \frac{\pd \log Z}{\pd \mu}
\notag \\
=&
\frac{1}{Z}\, \int_{-\pi}^{\pi} \prod_{n=1}^L \frac{dA_n}{2\pi}
\rme^{-S_b(A)}  \det D[A;\mu] \, n[A;\mu]
\notag \\
=&
\frac{I_1(\beta)^L\,  \sinh L\mu}
        { I_1(\beta)^L \cosh L \mu + I_{0}(\beta)^L \cosh L\hat m }
\end{align}
with
\begin{align}
n[A; \mu] =&
\frac{\sinh (L\mu+\rmi s) }{\det D[A;\mu]}
\, ,
\end{align}
and 
\begin{align}
\langle \sigma \rangle   = &  \frac{1}{L}    \frac{\pd \log Z}{\pd m}
\notag \\
=&
\frac{1}{Z}\, \int_{-\pi}^{\pi} \prod_{n=1}^L \frac{dA_n}{2\pi}
   \rme^{-S_b[A]}  \det D[A;\mu] \, \sigma[A;\mu]
\notag \\
=&
\frac{I_0(\beta)^L \, \sinh L \hat m}
       { I_L(\beta)^L \cosh L \mu + I_{0}(\beta)^L \cosh L\hat m }
\frac{1}{\cosh \hat m}
\end{align}
with
\begin{align}
  \sigma[A;  \mu]  =& 
  \frac{\sinh L\hat m}{\det D[A;\mu]}\,
  \frac{1}{ \cosh \hat m} 
\, .
\end{align}
At zero chemical potential $\mu=0$,
the fermion number density vanishes $\langle n \rangle =0$
and the scalar density is nonzero $\langle \sigma \rangle \ne 0$.
These densities show crossover-transition behavior
as the chemical potential $\mu$ is increased
at fixed $L$, or temperature $T=1/L$.
In the zero temperature limit $L\to \infty$,
the crossover behavior changes to a first-order transition
at a critical value of $\mu_c$.

\subsection{Uniform-field model}
For studying analytic properties of the CL sampling
in the complexified configuration space,
it is instructive to introduce a single-variable model which can be deduced
from the (0+1) dimensional Thirring model by setting uniform-field
condition, $ x \equiv A_n$ for all $n=1, \cdots, L$.
Then the bosonic action reduces to $S_b(x)=\beta  L  (1- \cos x )$
and the ``determinant'' term becomes
\begin{align}
  \det D_0(x;  \mu)=&  \cosh L( \mu + \rmi x)  + \cosh L\hat m
  \, .
\end{align}
The partition function of this model is given as
\begin{align}
Z_0 =& 
\int_{-\pi}^{\pi} \frac{dx}{2\pi} \, \rme^{-S_b(x)} \det D_0(x; \mu)
\notag \\
=&
 \rme^{-\beta L}
\left [ I_L(\beta L) \cosh L \mu + I_{\,0}(\beta L) \cosh L\hat m \right ]
.
\label{eq:Z0}
\end{align}
The fermion number density and scalar density are defined
in the same way as before:
\begin{align}
  \langle n \rangle
=&
\frac{1}{Z_0}\, \int_{-\pi}^{\pi} \frac{dx}{2\pi}   \rme^{-S_b(x)}  \det D_0(x;\mu) \, n(x;\mu)
\notag \\
=&
\frac{I_L(\beta L)\, \sinh L\mu}
        { I_L(\beta L) \cosh L \mu + I_{\,0}(\beta L) \cosh L\hat m }
\end{align}
with
\begin{align}
n(x; \mu) =&
\frac{\sinh L(\mu+\rmi x) }{\det D_0(x;\mu)}
\, ,
\label{eq:n0}
\end{align}
and
\begin{align}
  \langle \sigma \rangle
=&
\frac{1}{Z_0}\, \int_{-\pi}^{\pi} \frac{dx}{2\pi}   \rme^{-S_b(x)}  \det D_0(x;\mu) \, \sigma(x;\mu)
\notag \\
=&
\frac{ I_0(\beta L)\, \sinh L \hat m  }
       { I_L(\beta L) \cosh L \mu + I_{\,0}(\beta L) \cosh L\hat m }
\frac{1}{\cosh \hat m}
\end{align}
with
\begin{align}
\sigma(x;  \mu)  =& \frac{\sinh L\hat m}{\det D_0(x;\mu)}\frac{1}{ \cosh \hat m}
\, .
\label{eq:sigma0}
\end{align}
The parameter $L$ is a remnant of the lattice size, with which
we can study the system size dependence in this uniform-field model.
We use this model for examining the analytic properties of the CL sampling
and the severity of the sign problem.

\subsection{CL simulation}

For the Thirring model in (0+1) dimensions, 
the CL evolution in the Langevin time $\tau$ with discrete step $\varepsilon$,
is written collectively
for the complexified variables $z_\ell$ ($\ell=1, \cdots, L$)
as
\begin{align}
\label{eq:CL}
z(\tau+\varepsilon) = z(\tau)+\varepsilon K(z(\tau))
+ \sqrt{\varepsilon}\, \eta(\tau)
\end{align}
with the drift force
\begin{align}
K_\ell(z)=-\frac{\partial S(z)}{\partial z_\ell}
=
- \beta \sin z_\ell + \rmi
\frac{ \sinh (L\mu+ {\rm i} s) }
     {\det D(s;\mu)}
\label{eq:K}
\end{align}
and with the real stochastic force $\eta_n(\tau)$ whose average satisfies
$\langle \eta_n (\tau) \rangle_\eta =0$ and 
  $\langle \eta_n(\tau) \eta_{n'}(\tau')\rangle_\eta =
2 \delta_{nn'}\delta_{\tau,\tau'}$.
The sum $s\equiv \sum_{n=1}^{L} z_n$ becomes complex-valued, too.
For the uniform-field model we have a single equation
with the same structure as above,
where the drift force reads
\begin{align}
K(z)=-\frac{dS(z)}{dz}
=
-L \beta \sin z + \rmi L
\frac{ \sinh L(\mu+ {\rm i} z) }
     {\det D_0(z;\mu)}
     .
     \label{eq:K0}
\end{align}
Notice that the zeros of the determinant give rise to
singularities of the drift force in both cases.

Expectation values of observables $O(x)$ are computed with
the ensemble distribution $P(x,y)$
sampled along with a CL trajectory as
\begin{align}
  \left < O \right >_{\rm CL} = 
   \int dx dy\, P(x,y) O(x+\rmi y)\sim
  \frac{1}{N_{\rm sample}}\sum_\tau O(z(\tau))
  \, .
\end{align}

\begin{figure}
\begin{center}
\includegraphics[width=0.47\textwidth]{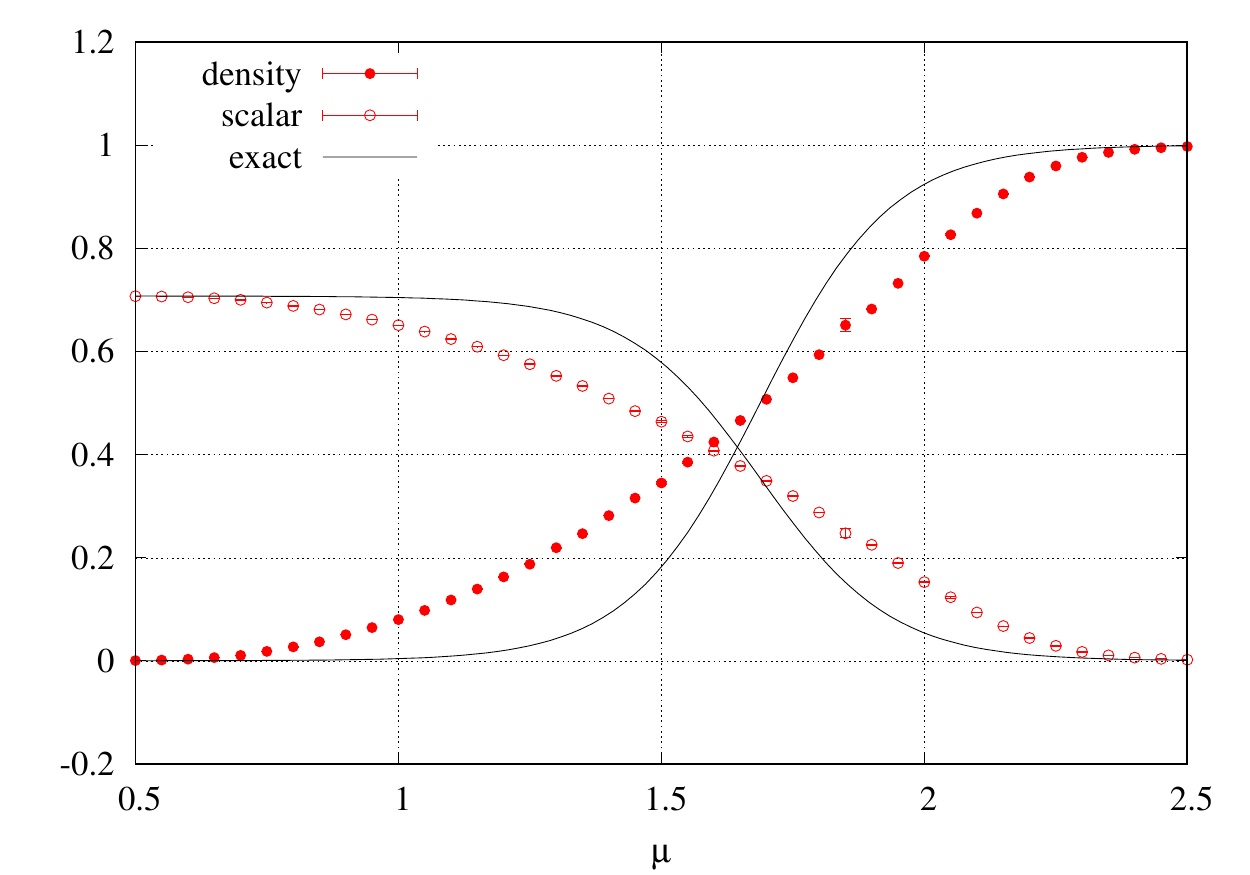}
\hfill
\includegraphics[width=0.47\textwidth]{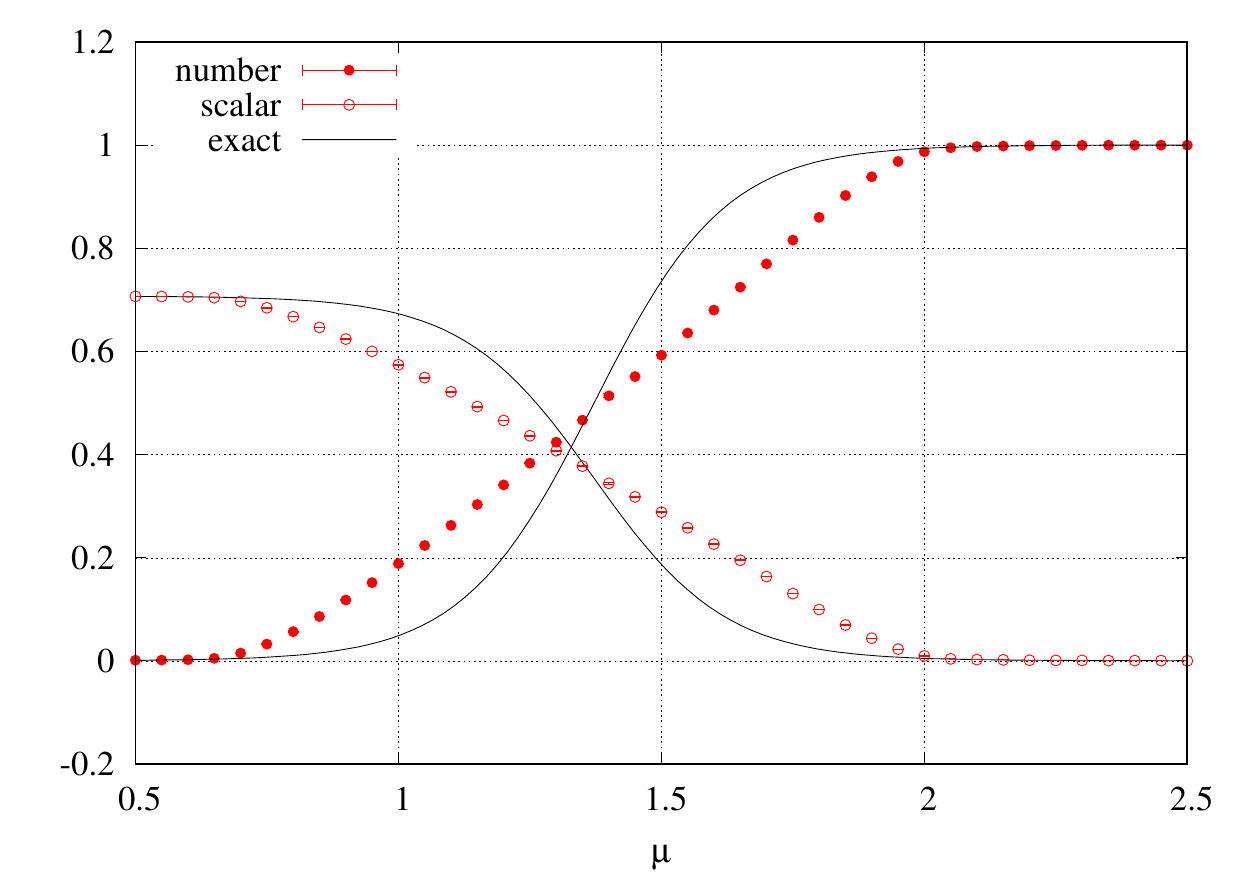}
\\
\vspace{10mm}
{}\hfill (a) \hspace{0.4\textwidth} (b) \hfill{~}
\end{center}
\caption{
  (a):  Fermion number $\langle  n\rangle$ (filled circle) and
  scalar $\langle \sigma \rangle$  (open circle) densities
  as a function of the chemical potential $\mu$
  in (0+1) dimensional Thirring model
for $L=8$, $\beta=1$ and $m=1$.
  (b):  the same plot for the uniform-field model.
}
\label{fig:CL-results}
\end{figure}


We have performed the CL simulation for these models
at finite~$\mu$.
We set the model parameters as $\beta=1$, $m=1$, $L=8$, and 
we choose $\varepsilon=10^{-5}$ for the Langevin time step.
After $10/\varepsilon$ steps for thermalization,
we evolve the equation by $10^9$ steps with sampling the
configurations in every $10^3$ steps.
When the drift magnitude $|K|$ becomes large such that
$\varepsilon |K|/L > \varepsilon_0=10^{-3}$,
we evolve the equation with a smaller step size $\varepsilon_v$
to keep $\varepsilon_v |K|/L < \varepsilon_0$
till $\varepsilon \le \sum \varepsilon_v$ is achieved.
Since the model is periodic in $\text{Re}z_\ell$, we restricted
$\text{Re}z_\ell \in [-\pi,\pi)$ in the simulations.

The simulation results for the fermion number density $\langle n \rangle$
and scalar condensate $\langle \sigma \rangle$ in the
Thirring model are
plotted as a function of $ \mu$ in Fig.~\ref{fig:CL-results}~(a).
The exact results (solid curves) clearly show smooth
crossover behavior in the region of $1 \lesssim \mu \lesssim 2.3$.
In contrast, the CL results for $\langle n \rangle$ and
$\langle \sigma \rangle$
deviate from the exact ones, showing milder dependence on
$\mu$ around the crossover region.
They agree with the exact results for small and large values
of $\mu$. A similar trend is also seen in the CL result
for the uniform-field model as shown in Fig.~\ref{fig:CL-results}~(b);
the exact solution shows crossover transition
in the region of $0.8 \lesssim \mu \lesssim 2.0$, while the CL result behaves
almost linearly with increasing $\mu$ in the corresponding region.

\subsection{CL ensembles, determinant zeros and thimbles}

\begin{figure}[ht]
\begin{center} 
\hfil
\includegraphics[width=0.35\textwidth, bb=0 0 360  360]{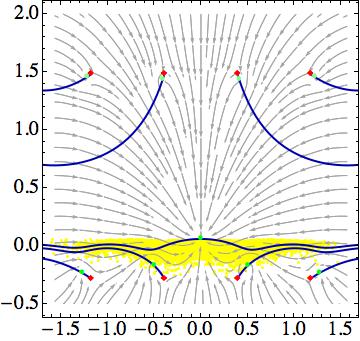}
\hfil
\includegraphics[width=0.35\textwidth, bb=0 0 360  360]{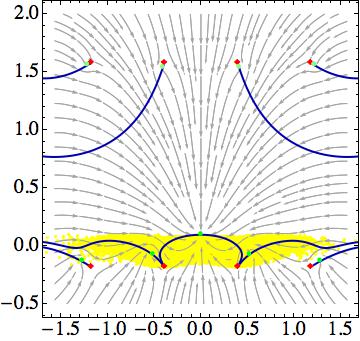}
\hfil
\\
{}\hfil (a) \hfil \hfil (b) \hfil{~}
\\
\includegraphics[width=0.35\textwidth, bb=0 0 360  360]{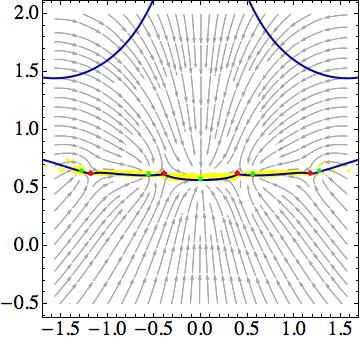}
\\
\hfil (c) \hfil {~}
\\
\hfil
\includegraphics[width=0.35\textwidth, bb=0 0 360  360]{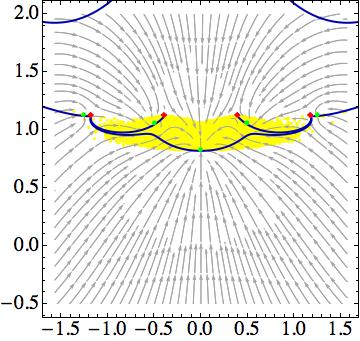}
\hfil
\includegraphics[width=0.35\textwidth, bb=0 0 360  360]{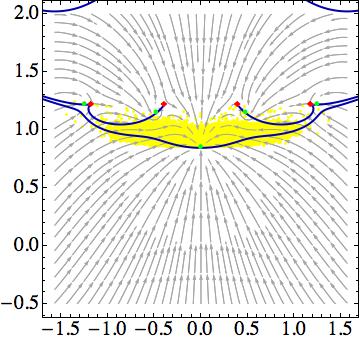}
\hfil
\\
{}\hfil (d) \hfil \hfil  (e) \hfil{~}
\end{center}
\caption{
  Scattering plot of the CLE sampling for $\mu=0.5$ (a), 0.6 (b)
  1.5 (c),  2.0 (d) and 2.1 (e)
  with $L=8$, $m=1$, $\beta=1$.
  Arrows depict the field of the drift force
  and  blue curves are thimbles.
  Red diamonds and green circles are
  singular and zero points of the field, respectively.
}
\label{fig:ScattPlots}
\end{figure}

It has been noticed for some time
that the CL method fails in crossover regions for
simple fermionic models with singular drift terms,
  such as the chiral random matrix model,
  the models with modified Rayleigh distributions,
Hubbard model, etc., in addition to the (0+1) dimensional
Thirring model
\cite{SanoLattice2011,Pawlowski:2013pje,Pawlowski:2013gag,
Mollgaard:2013qra,
Mollgaard:2014mga,
Hayata:2015lzj}.
In order to understand the situation clearly,
we present in Fig.~\ref{fig:ScattPlots}
the scatter plots of the CL samples in the uniform field model
for $\mu=0.5, 0.6, 1.5, 2.0$ and $2.1$ with $L=8, \beta=1, m=1$.
We collected $10^5$ samples and show them with yellow dots
in the complex $z$ plane.
In the same figure,
we also display the drift flow field (\ref{eq:K0})
with gray arrows,
and the zeros of the determinant in red diamonds,
where the drift become singular.  
Green circles denote the fixed points $z_{\rm c}$
of the drift force field.

Blue curves are the steepest descent paths ${\cal J}_{z_{\rm c}}$
for $\rme^{-S}$,
or the thimbles in Lefschetz theory.
The thimbles are defined by the union of integral curves of
the anti-holomorphic flow
\begin{align}
  \frac{d z}{d\tau} = \overline{\frac{\partial S[z]}{\partial z}}
\end{align}
with the boundary condition, $\lim_{\tau \to -\infty}z(\tau) \to z_{\rm c}$.
It is known that a set of thimbles $\{ {\cal J}_{z_{\rm c}}\}$
becomes equivalent to the original integration contour:
\begin{align}
  Z=\int_{-\pi}^\pi dx \rme^{-S} = \int_{\{\cal J\}} dz \rme^{-S}
  .
\end{align}
One important property of the anti-holomorphic flow is that ${\rm Re}S$
is monotonically increasing with ${\rm Im}S$ kept constant
along the flow, especially on a thimble.
This property motivated us to perform Monte Carlo simulations on
thimbles\cite{Cristoforetti:2012su,Cristoforetti:2013wha,Fujii:2013sra}.

This anti-holomorphic flow field is obtained from the drift force field by
flipping the sign of the real component.
Under this transformation,
zero points $z_{\rm zero}$ and critical points $z_{\rm c}$,
which are saddle points and attractive (or repelling) points in CL method,
respectively become attractive points
\footnote{
  Since we assume that $S$ is bounded below, we don't have repelling points here.
}
and saddle points in Lefschetz theory.
Since the total set of thimbles form a skeleton graph without loops,
there must be an associated critical point $z_{\rm c}$ for each 
zero point $z_{\rm zero}$, which provides an endpoint for a thimble ${\cal J}_{z_{\rm c}}$.

In Fig.~\ref{fig:ScattPlots}, we first notice 
that the critical point $z_{0}$ on the imaginary axis is an attractive
point of the drift force field for all the values of $\mu$ studied here,
and it is always included in the CL ensemble.
Indeed, thimble ${\cal J}_{z_0}$ is always the most dominating one
in the thimble integration in this model.
At $\mu=0.5$ (Fig.~\ref{fig:ScattPlots} (a)),
the distribution of the CL samples
is elongated along the original integration
contour, the real axis $[-\pi,\pi)$, having small
extent in the imaginary direction.
It also has a good overlap with the thimble ${\cal J}_{z_0}$
emerging from the critical point ${z_0}$.
We note that the distribution apparently has no overlap
with the vicinity of any $z_{\rm zero}$.

At a slightly larger chemical potential $\mu=0.6$
(Fig.~\ref{fig:ScattPlots} (b)), however,
the distribution gets expanded in the imaginary direction
to have an overlap with the vicinity of some zeros $z_{\rm zero}$
of the determinant.
This situation is understood as follows:
With increasing $\mu$, the zeros  $\{z_{\rm zero}\}$ 
move upward together with their associated critical points
\footnote{
  See Appendix A for the explicit expressions for $z_{\rm zero}$'s.
},
and a reconnection of the thimbles (the Stokes phenomenon)
occurs between $\mu=0.5$ and 0.6.
At $\mu=0.6$ the thimble ${\cal J}_{z_0}$ terminates at
the two zeros closest to the origin,
and the integration contour now consists of multi thimbles.
A change of the thimble structure must accompany
a change of the drift flow pattern,
because the anti-holomorphic flow defining the thimbles
is obtained just by flipping the sign of the real component of
the drift flow \ref{eq:K0} of CL method.

We show a representative configuration of a CL ensemble,
drift field and thimbles at $\mu=1.5$ in Fig.~\ref{fig:ScattPlots}~(c),
where the thimbles are connected with each other at $z_{\rm zero}$'s,
and the CL samples are distributed closely to the thimbles
\footnote{
  It is known that 
  there is other models where a critical point of a relevant thimble
  becomes repulsive and not sampled in the CL sampling.
}.
Until $\mu \sim 2.0$ (Fig.~\ref{fig:ScattPlots} (d)),
the CL sampling includes the vicinities
of the zeros, and exactly in this region of $\mu$,
the CL results converge to the incorrect solution.
Above this value of the chemical potential,
as is seen in Fig.~\ref{fig:ScattPlots} (e),
the CL samples are distributed away from any zeros
and localized around the single thimble ${\cal J}_{z_0}$
covering practically the whole integration contour.
In this case, the CL simulation gives the the correct result
(see Fig.~\ref{fig:CL-results} (b)).

\begin{figure}[tb]
\begin{center}
  \includegraphics[width=0.6\textwidth,bb=0 0 461 346]{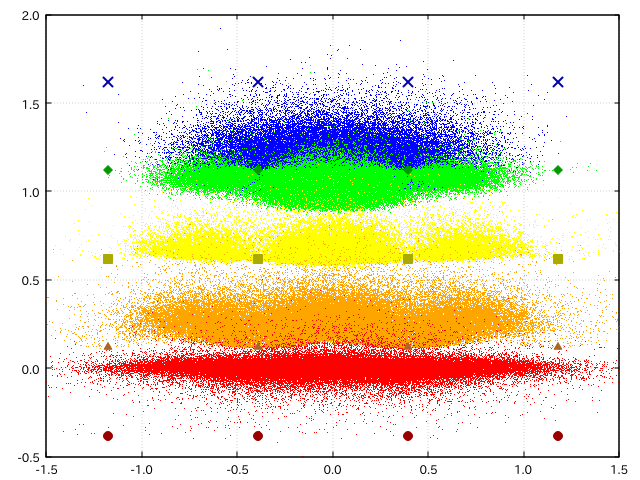}
\end{center}
\vspace{1cm}
\caption{Scatter plot of the CL samples of 0+1 dim model with
  $L=8$, $\beta=1$ in
  complex $s/L=\sum_{n=1}^L z_n/L$ space for $\mu=0.5$,
  1, 1.5, 2, and 2.5 from bottom to top.
  Singular points of the drift force are shown with circles,
  triangles, squares, diamonds, and crosses for respective values of
  $\mu$.
  For $\mu=1.0, 1.5$ and 2.0, these singular points locate
  at the edges of scatter distributions.
  Note that the distribution for $\mu=2.5$ is almost
  oval-shaped, but part of it is hidden below that for $\mu=2.0$.}
\label{fig:ScattPlot0+1}
\end{figure}

Importance of multi-thimble contributions in
the transition region is emphasized in
Refs.~\cite{Tanizaki:2015rda,Fujii:2015bua};
each thimble ${\cal J}_{z_c}$
has a complex phase of $\text{Im}S(z_c)$,
and destructive interference among multi-thimble contributions
is crucial for realizing the Silver-Blaze
phenomenon\cite{Cohen:2003kd},
i.e.,
no response of the system to $\mu$ below a certain critical value
at zero temperature.

Because the thimbles in the set which is equivalent to the original
integration contour are connected at the determinant zeros
and because the CL samples are well localized around 
the thimbles in this model, 
it seems almost inevitable that the CL samples
the neighborhoods of the drift singularities (determinant zeros).
In Fig.~\ref{fig:ScattPlots} (c), the critical points and zero points
are aligned almost in a straight-line parallel to the real axis, and then
these critical points become attractive nodes and
these zeros are saddle points in the drift flow field.
We also notice that in the thimble integration one needs to
include correctly the complex phases from $\text{Im}S(z_c)$ and
the Jacobian in the change of the variables,
while
in the CL method observables are computed 
as a simple average of the samples $P(x,y)$
distributed around the thimbles
in the complex $z=x+\rmi y$ plane
without any additional phase factors.

Outside the crossover region, the integration contour is
dominated solely by a single thimble,
and relative phases among the thimbles become irrelevant.
In the CL approach, this implies that
the vicinities of the determinant zeros are not sampled
so that the simulations give the correct results.

In Fig.~\ref{fig:ScattPlot0+1}, we show a scatter plot of
the (0+1) dimensional model in complex $s/L=\sum_{n=1}^L z_n /L$ space,
for $\mu=0.5, 1, 1.5, 2$ and $2.5$
with  $L=8$, $\beta=1$, and $m=1$ in the complex $s/L$ plane.
The zero manifold, where $K=0$, projected on this plane becomes
a set of points, which we denote with circles, triangles, squares,
diamonds, and crosses for respective values of $\mu$
in Fig.~\ref{fig:ScattPlot0+1}.

We confirm that the CL ensembles overlap with the neighborhoods
of the determinant zeros for $\mu$ in the crossover region
($\mu=1, 1.5, 2$).
We also note that the samples are scattered wider in imaginary
direction than in the uniform-field model.


\begin{figure}
  \begin{center}
    \hfil
    \includegraphics[width=0.45\textwidth]{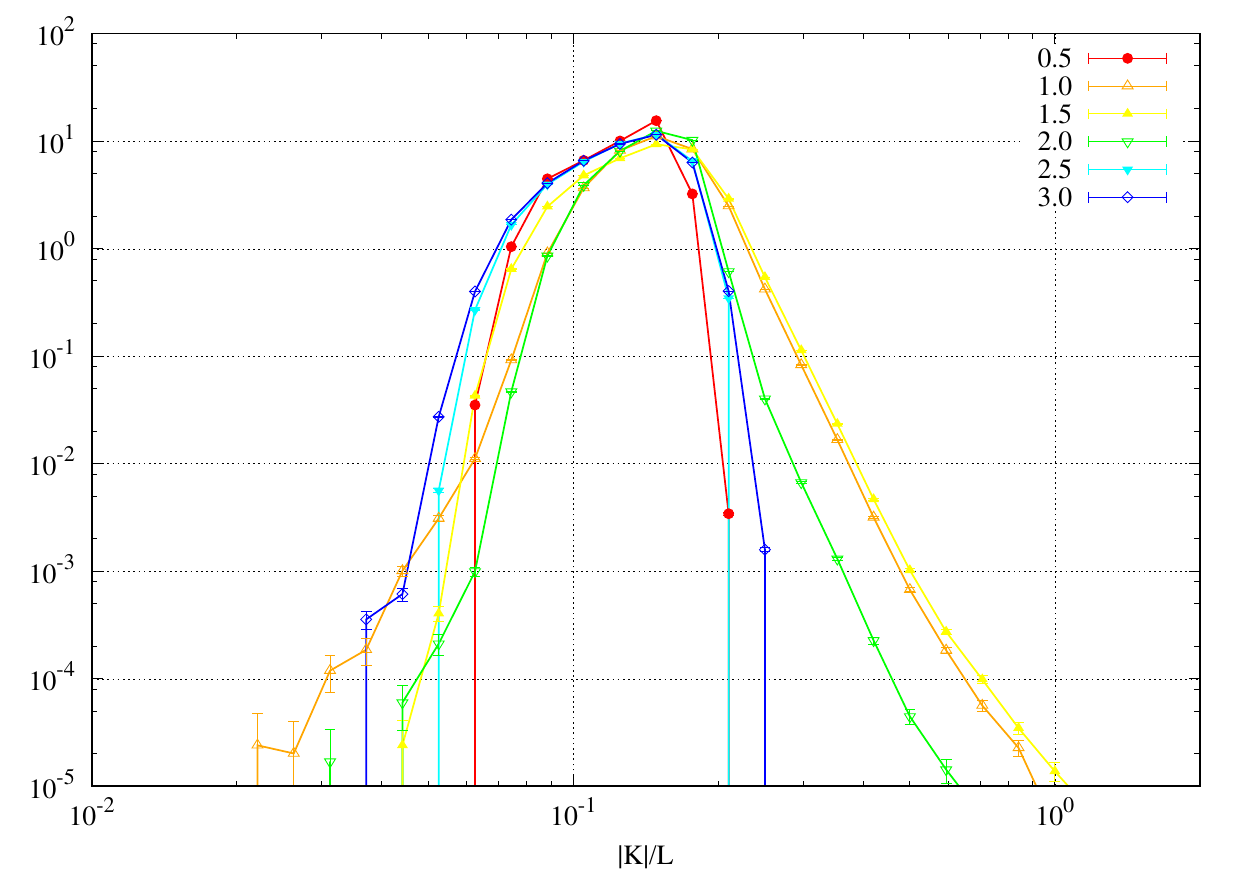}
    \hfill
    \includegraphics[width=0.45\textwidth]{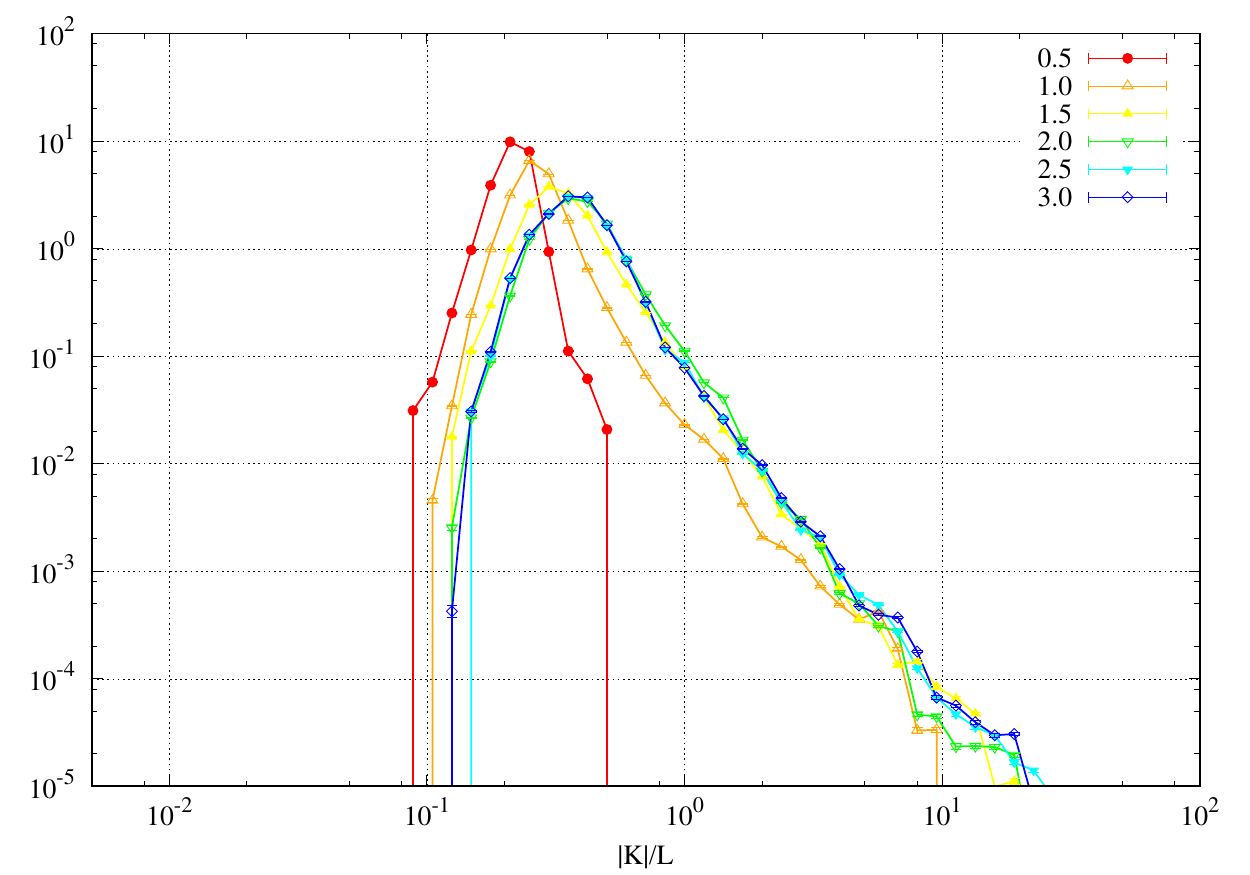}
\\
\vspace{10mm}
{}\hfill (a) \hspace{0.4\textwidth} (b) \hfill{~}
  \end{center}
  \caption{
  (a):  Histogram of magnitude of the drift term $|K(z)|$
    in the uniform-field model for $\mu=0.5, 1, 1.5, 2, 2.5$
    and $3$ (with $10^5$ samples).
  (b): The same plot for 
    in (0+1)-dim.\ model
  for typical values of $\mu=0.5$, 1.0, 1.5,2.0, 2.5 and 3.0  
  }
\label{fig:K-histogram}
  \end{figure}
  \begin{figure}
    \begin{center}
\includegraphics[width=0.45\textwidth]{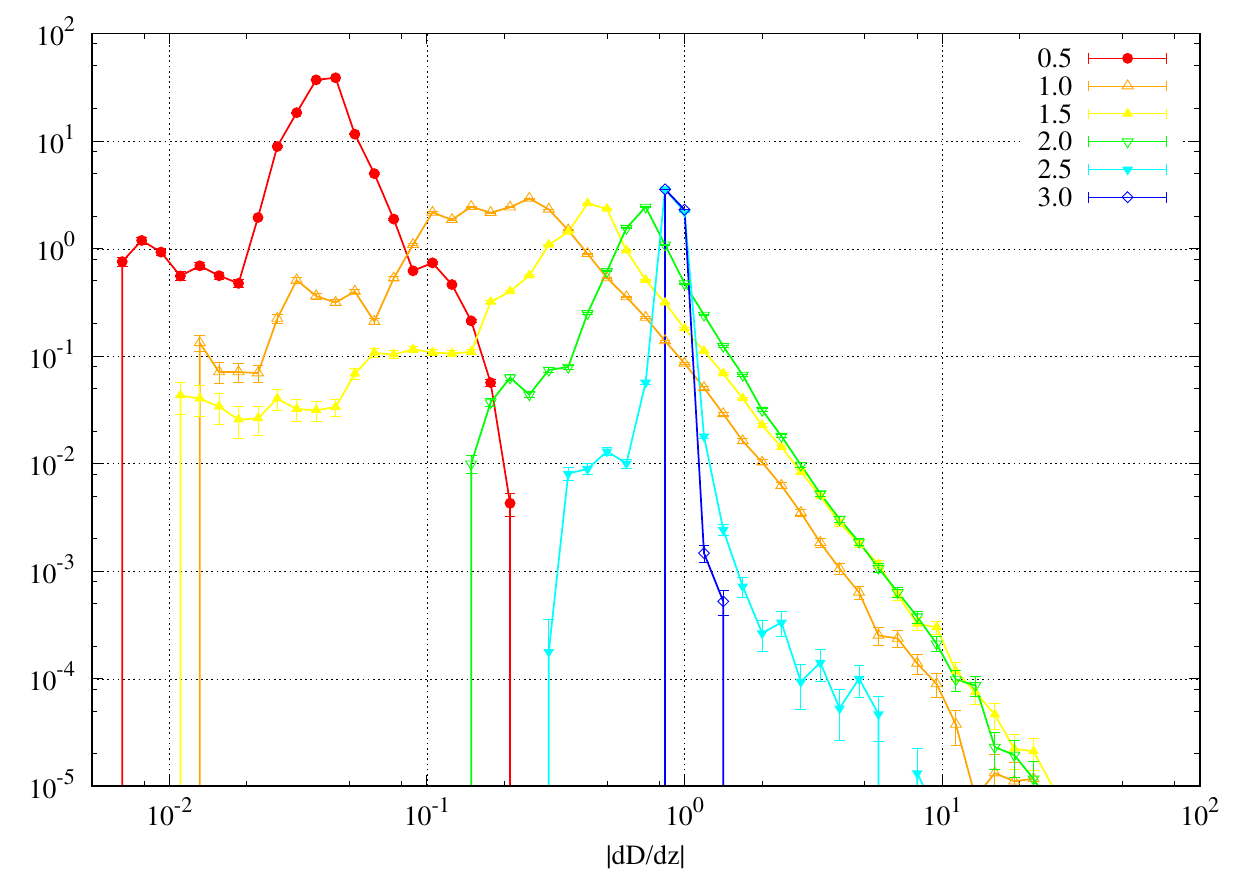}
\hfill
\includegraphics[width=0.45\textwidth]{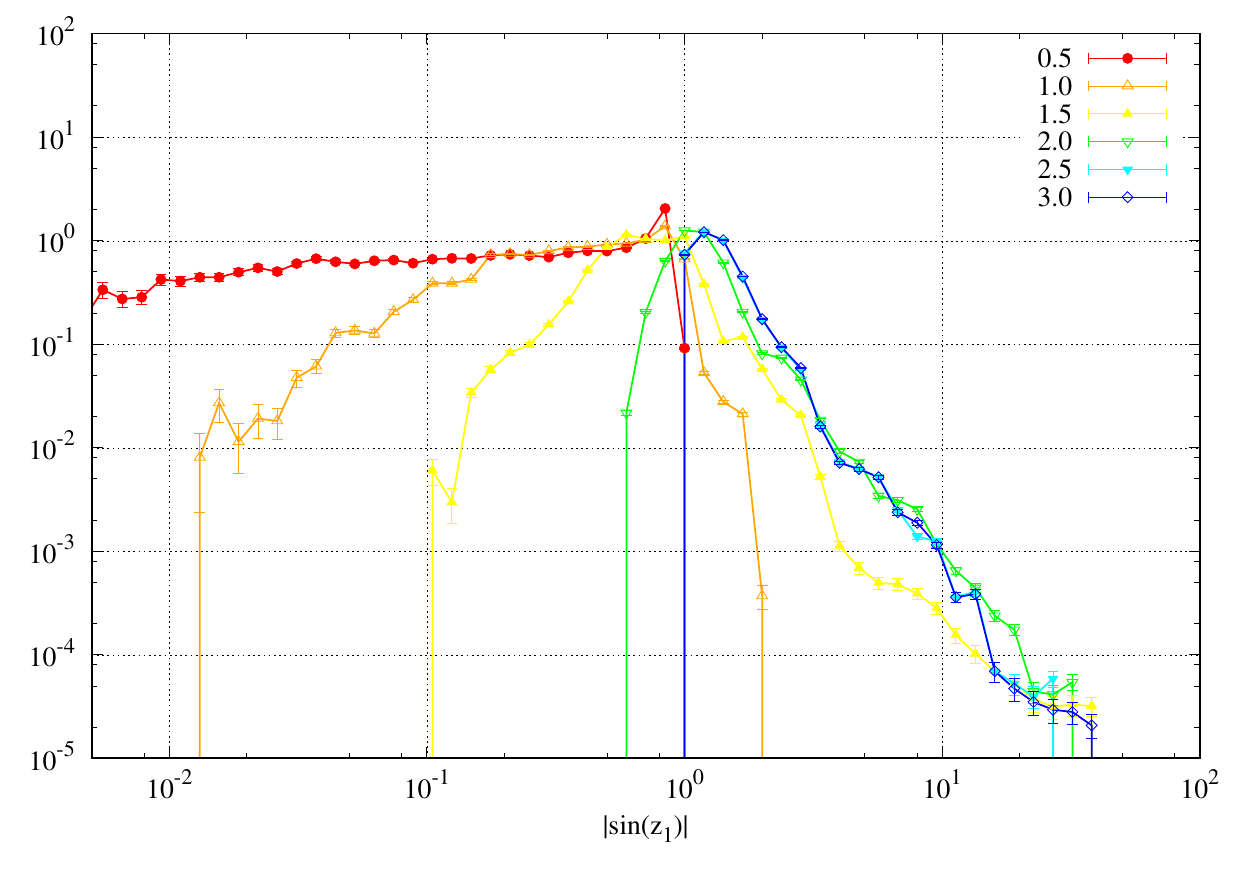}
\\
\vspace{10mm}
{}\hfill (a) \hspace{0.4\textwidth} (b) \hfill{~}
\end{center}
\vspace{4mm}
\caption{
Histograms of $|\frac{d}{ds} \log \det D| $ (a) and
$|\sin(z_1)|$ (b) in (0+1)-dim.\ model for $L=8$, $\beta=1$ and $m=1$.
}
\label{fig:dDdz-histogram}
\end{figure}
  
\subsection{Histogram of drift force}
Recently authors of Ref.~\cite{Nagata:2016vkn}
claimed that an exponential fall-off of the probability of
the drift term at large magnitude
is the necessary and sufficient condition for the correctness of
the CL results. Indeed, the histogram of $|K|$
is very helpful to quantify
the contributions from the vicinities of the zeros in the CL simulations.

We make histograms of the drift force $|K|$
of the uniform-field model with $L=8$, $\beta=1$ and $m=1$
for $\mu=0.5, 1, 1.5, 2, 2.5$ and $3$
in Fig.~\ref{fig:K-histogram}~(a)
by binning the drift magnitude as
$a^{n-1} \leq |K|/L < a^n$ (we chose here $a=2^{1/4}$ and $n$ integer).
Outside the crossover region, $\mu=0.5, 2.5$ and $3$,
we find that
the histograms fall off very sharply and become vacant at large $|K|$.
But it develops a power-law tail for $\mu=1, 1.5, 2$ in the crossover region,
which clearly show the fact that the CL sampling hits the vicinities of
the determinant zeros.
This result
is consistent with the argument of Ref.~\cite{Nagata:2016vkn}.

In Fig.~\ref{fig:K-histogram}~(b),
we make histograms of $|K|$ 
of the (0+1) dimensional Thirring model with $L$=8, $\beta=1$ and $m=1$
for $\mu=0.5$, 1.0, 1.5, 2.0, 2.5, 3.0.
At small $\mu=0.5$, for which the CL simulation is successful,
the histogram is well localized around $|K|/L\sim 0.2$.
But we find that
a power-law-like distribution of $|K|$ appears for $\mu \gtrsim 1.0$,
which may be caused by the singularities of the force at
the determinant zeros.
From Fig.~\ref{fig:CL-results} (a),  however,
the value $\mu=3$ is well outside the crossover region
and the CL simulation successfully reproduce the correct result.
Moreover,
from Fig.~\ref{fig:ScattPlot0+1} the vicininities
of the zeros are little sampled at this value of $\mu$,
in spite of the power-low tail appearing in the histogram.
.

To measure the contributions from the vicinities of
the drift singularities, we make histograms of
$|\tfrac{d}{ds}\log \det D(s)|$ in Fig.~\ref{fig:dDdz-histogram}~(a).
We find that the histogram is empty at large
$|\tfrac{d}{ds}\log \det D(s)|$
for $\mu=0.5$ and 3.0, {\it i.e.,} for $\mu$ outside the crossover region,
while the power-law tail appears for $\mu$ in the crossover. 
It is concluded, therefore, 
that the power-law-tail contribution from the neighborhoods
of the drift singularities exists only in the crossover region
to make the CL method dubious there.

Then the power-law tail at large $|K|$ in the
histogram Fig.~\ref{fig:K-histogram}~(a)
for large $\mu$
must come from the first term $-\beta \sin(z_\ell)$
in the drift force, (\ref{eq:K}).
This is indeed the case as can be confirmed by looking at 
the histogram of $|\sin(z_1)|$, for example,
in Fig.~\ref{fig:dDdz-histogram}~(b).
We see that the probability of large $|\sin(z_1)|$
develops as $\mu$ increases.

\begin{figure}[tb]
  \centerline{
    \includegraphics[width=0.5\textwidth, bb=0 0 461 346]
                  {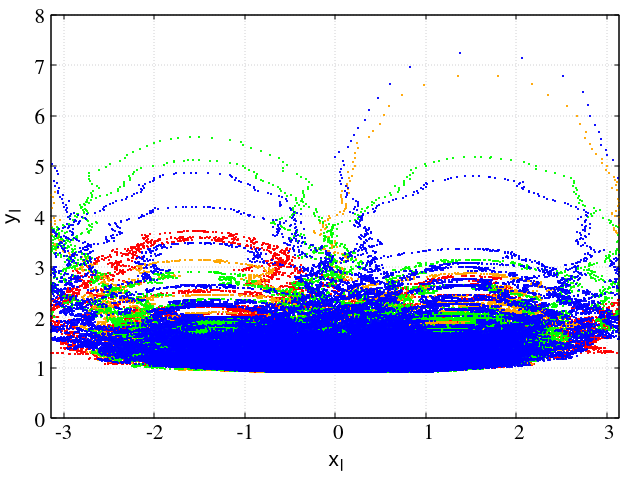}
  }
  \caption{Example of trajectory components, $z_{1,2,3,4}$,
    for $L=8$, $\beta=1$ and $m=1$ sampled in every $10^2$ steps
    with time step $\epsilon=10^{-5}$.
    Other components $z_{5,6,7,8}$ behave similarly.}
\label{fig:trajectory_z_ell}
\end{figure}

This power-law tail of the $|K|$-distribution for large $\mu$ (=3)
appears by fluctuations of $z_\ell$ ($\ell=1, 2, 3, \cdots, L$)
in imaginary direction.
In the (0+1) dimensional model,
with $s=\sum_{n=1}^L z_n$ localized around the critical point $z_0$
(Fig.~\ref{fig:ScattPlot0+1}),
a component $z_\ell$ can first fluctuate in real direction
to the region $z_\ell \sim \pm \pi + \rmi y_\ell$ with $y_\ell>0$.
Then a trajectory rides on a drift flow to imaginary direction, 
which is caused by $-\beta \sin z_\ell$ term in the flow,
(\ref{eq:K}),
and the trajectory makes an excursion in the imaginary direction
which then
comes back to the region near $z_0$ (See Fig.~\ref{fig:trajectory_z_ell}).
We have confirmed that the histogram of $y_\ell$ integrated over other
variables is consistent with an {\it exponentially decaying} distribution
$\propto \rme^{-c y_\ell}$  with $c\sim 2.6$
for $y_\ell \gtrsim 2$.
When plotted with respect to $|\sin z_\ell| \sim \tfrac{1}{2} \rme^{y_\ell}$,
the histogram decays with power law $|\sin(z_\ell)|^{-(c+1)}$
in large $y_\ell$ region, which is seen in Fig.~\ref{fig:dDdz-histogram}~(b).

\subsection{Short summary}

We have seen that the CL simulation converges to an incorrect solution
within the crossover region of the model,
where CL samples the neighborhoods of the determinant zeros,
$z_{\rm zero}$, as seen in the histograms
of the drift force and also in the scatter plots.
This is the very situation where CL method becomes dubious
as is argued in Refs.~\cite{Nagata:2016vkn,Aarts:2017vrv}.
We have also seen that
it is almost inevitable for CL simulations to sample the vicinities of
these zeros $z_{\rm zero}$'s from the observation of the drift flow pattern
in the crossover region, where contributions from multi-thimble connected
with each other via these zeros become important in the thimble integration.

Outside the crossover region, on the other hand,
the CL simulation successfully reproduces the correct solution.
In this region, the thimble structure
in the complexified configuration space is so simple that
a single thimble becomes practically equivalent
to the original integration domain,
and the CL ensemble is well localized around this single thimble
without sampling the vicinity of any determinant zeros $z_{\rm zero}$'s.

In order to apply the CL simulation in the crossover region,
therefore, we need methods which avoid sampling the vicinity of $z_{\rm zero}$'s.
One possible approach to overcome this situation
is the so-called re-weighting method, in which
one evaluates the observables in the crossover region
by using the ensemble generated in other parameter region
where the CL method gives the correct results\cite{Bloch:2017ods}.
Another approach is to express the observables with a deformed (or modified) model
for which the CL simulation is successful without sampling the vicinity
of its zeros\cite{Tsutsui:2015tua}.

We examine these two approaches one by one in the next two sections.

\section{Reweighting method at finite $\mu$}

In order to compute the observables 
correctly in the crossover region,
one can try the reweighting method
with CL ensembles in
the parameter space ($\beta, m, \mu$) where CL method works well
\cite{Bloch:2017ods}.
We have observed that the CL simulation is successful
for $\mu$ outside the crossover region,
where a single thimble practically covers the original integration contour
and the CL ensembles are distributed along this thimble.
Based on this observation,
we study here the reweighting method
with the CL ensembles generated
at a large reference chemical potential,
which we call $\nu$.

In this approach, we recast the expectation value of an observable $O$ as
\begin{align}
  \langle O \rangle_\mu
  \equiv&
  \frac{\int_{-\pi}^{\pi} \frac{dx}{2\pi} \rme^{-S_b(x)} \det D(x; \mu)O(x)}
       {\int_{-\pi}^{\pi} \frac{dx}{2\pi} \rme^{-S_b(x)} \det D(x; \mu)}
      = \frac{\langle\frac{ \det D(\mu)}{ \det D(\nu)} O \rangle _\nu}
       {\langle\frac{\det D(\mu)}{\det D(\nu)}  \rangle _\nu}
       \, ,
       \label{eq:RWformula}
\end{align}
where the subscript $\langle ~ \rangle_\mu$
indicates that the expectation value is
evaluated with the weight $\exp(-S_b(x)) \det D(x;\mu)$
and similarly for $\langle ~ \rangle_\nu$.
In usual reweighting, one introduces a real weight function
so that one can perform Monte Carlo sampling.
For our model we can choose
$\exp(-S_b(x)) \det D(x;0)$ or
$\exp(-S_b(x)) |\det D(x;\nu)|$
as the weight function.
But in the context of CL simulations,
one may take the system with large chemical potential $\nu$
outside the crossover region as the reference system,
and perform the CL simulation to evaluate the observables
using Eq.~(\ref{eq:RWformula}).
The reweighting method with the CL method
is expected to work
as long as $\nu$ is outside the crossover region
but the reference ensembles still have sufficient overlap
with the physical one, as discussed below.

\subsection{Reweighting factor}

Severity of the sign problem in this approach appears in the evaluation
of the reweighting factor with the reference ensemble
\begin{align}
\left \langle  \frac{\det D( \mu)  }  { \det D( \nu) } \right \rangle_\nu
\equiv &
\frac
    {\int _{-\pi}^\pi dx\,  \rme^{-S_b(x)}  \det D(x;  \nu)
    \frac{\det D(x;  \mu)  }{\det D(x;  \nu) }   }
{\int _{-\pi}^\pi dx\,  \rme^{-S_b(x)}  \det D(x;  \nu)  }
\hspace{1.5cm}
\left (=  \frac{Z(\mu)}{Z(\nu)} \right )
\notag \\
= &
\frac {\int_{\sum {\cal J}} dz\,  \rme^{-S_b(z)} \det D(z; \nu)
      \frac{\det D(z; \mu)  }  { \det D(z; \nu) }   }
      {\int _{\sum {\cal J}} dz\, \rme^{-S_b(z)} \det D(z; \nu)  }
      .
      \label{eq:RWfactor}
\end{align}
In the second line, 
we have changed the integration contour
from the real axis to a set of the thimbles
for the function $\rme^{-S_b(z)} \det D(z; \nu)$,
because the integrands are holomorphic.
In the CL simulation with large $\nu$,
the ensembles are localized along a single thimble which
covers the dominant part of the integration contour in the complex plane.
This suggests that the integration with the CL method is closely
related to the integration along the thimble ${\cal J}$,
{\it i.e.}, the second expression of eq.~(\ref{eq:RWfactor}).

It is noteworthy that the reweighting factor is expressed with
the average phase fluctuation factor 
Eq.~(\ref{eq:PhaseFactor2}) as
\begin{align}
\left \langle \frac{ \det D( \mu)  }{ \det D( \nu) }  \right \rangle_\nu
=
\left <\rme^{i\varphi}\right >_{\nu, {\rm p.q.}}
\cdot
\left <\left |\frac{\det D(\mu)}{\det D(\nu)}\right | \right >_\nu
\, ,
\label{eq:Reweight-Phase}
\end{align}
where the average phase fluactuation
is defined as the integral along the thimbles:
\begin{align}
\left \langle  \rme^{\rm i \varphi}  \right \rangle_{\nu,{\rm p.q.}}
\equiv &
\int _{\sum {\cal J}}  dz\, \rme^{-S_b(z)} \det  D( \nu)  \,  \frac{\det D( \mu)  }  { \det D( \nu) }  
\Big /
\int _{\sum {\cal J}}  dz\, \rme^{-S_b(x)} \det D( \nu)    \left |   \frac{\det D( \mu)  }  {\det D( \nu) }  \right |   
\, .
\label{eq:PhaseFactor2}
\end{align}
This clearly shows that the average phase factor (\ref{eq:PhaseFactor2})
is a good measure of the severity of the sign problem
in evaluation of the reweighting factor.
Note that the denominator of Eq.~(\ref{eq:PhaseFactor2})
is now an integral of a non-holomorphic function,
and that the average phase factor
(\ref{eq:PhaseFactor2}) depends on the choice of the integration path.
By setting $\nu=0$,
the integration path reduces to the real axis $[-\pi,\pi)$ and
  the expression (\ref{eq:PhaseFactor2}) coincides with
  the usual phase factor
\begin{align}
\left \langle \rme^{\rmi \phi} \right \rangle_{\text{p.q.} }\equiv
\frac{ \int_{-\pi}^\pi dx  \, \rme^{-S_b(x)}\, \det D(x;  \mu)  }
       { \int_{-\pi}^\pi dx  \, \rme^{-S_b(x)} \, |\det D(x;  \mu)|  }
\label{eq:PhaseFactor1}
\end{align}
because $\det D(x;\nu)$ is real positive there.

\begin{figure}
\begin{center}
\includegraphics[width=0.5\textwidth]{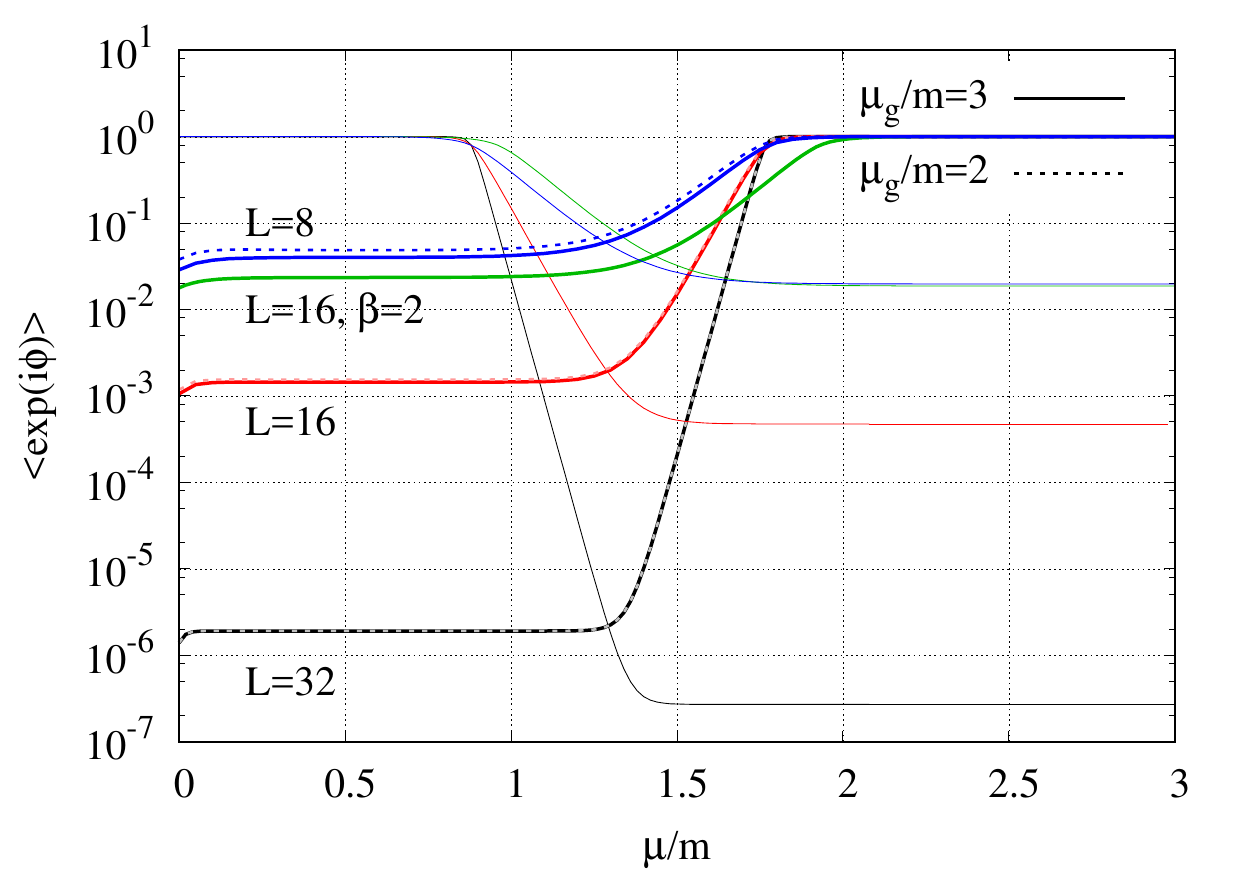}
\end{center}
\vspace{5mm}
\caption{Average of the phase factors
  $\langle \rme^{\rmi \phi} \rangle_{\rm p.q.}$ and
 $\langle \rme^{\rmi \varphi} \rangle_{\nu,{\rm p.q.}}$ as a function of $\mu/m$ 
  for $L=8,16,32$ with $\beta=1, m=1$, and
  for $L=16$ with $\beta=2, m=1/2$.
  The reference chemical potential $\nu/m$ is set to
  3 (solid) and 2 (dashed).
} 
\label{fig:Phase}
\end{figure}

We plot in Fig.~\ref{fig:Phase}
the averaged phase factors,  $\langle \rme^{\rmi \phi} \rangle_\text{p.q}$ and
$\langle \rme^{\rmi \varphi} \rangle_{\nu,{\rm p.q.}}$
as a function $\mu/m$ with fixed $\nu =2$ (dashed) and 3 (solid)
for $L=8, 16$ and $32$ in the uniform-field model.
The phase average $\langle \rme^{\rmi \phi} \rangle_\text{p.q.}$ with
$\mu=0$ is
unity at small $\mu$ and
starts to decrease around $\mu \sim 0.8$, which
is close to the threshold value for the failure of the simple CL simulation. 
It continues to decrease until around $\mu \sim 1.5$
and becomes constant for larger $\mu$.
But the value of the phase average
$\langle \rme^{\rmi \phi} \rangle_{{\rm p.q.}}$
at large $\mu$ is suppressed exponentially in~$L$;
it  becomes as small as $10^{-2}$ for $L=8$ at large $\mu$,
and in order to get signal at large $\mu$,
one needs at least $O(10^{4})$ samples in statistical sampling.
For $L=16$ the number of required samples
is likely to become more than $O(10^{6})$.
For $L=32$ it is more than $O(10^{13})$.

In contrast, the phase average
$\langle \rme^{\rmi \varphi } \rangle_{\nu,{\rm p.q.}}$
evaluated along the thimbles
is almost unity on the large $\mu$ side.
This can be understood by noting the fact that
the location of the dominant thimble ${\cal J}_{z_0}$
becomes less sensitive to the value of $\mu$
when $\mu$ is set to a larger value above the crossover region.
One can show that for large $\mu$
the critical point $z_0$ appears
at $\sin z_0 \sim {\rmi} / {\beta}$
on the imaginary axis, independently of $\mu$.
Then the thimbles ${\cal J}_{z_0}$ for large $\mu$ and $\nu$
locate very close to each other, on which the integrands are real
positive up to the Jacobian,
and therefore the phase factor
in eq.~(\ref{eq:PhaseFactor2}) becomes close to unity.%
\footnote{In CL simulations
  the integration measure is replaced with the ensemble average,
and the ensembles generated
by the CL evolution with these $\mu$ and $\nu$ are
closely overlapped with each other.}
At small $ \mu \lesssim \hat m$, on the other hand,
the phase average $\langle \rme^{i\varphi} \rangle_{\nu,{\rm p.q.}}$
becomes small because the integrand of $Z(\mu)$
is now complex-valued and very oscillatory along the dominant thimble
${\cal J}_{z_0}$ of $Z(\nu)$ in the complex $z$ plane.
We notice here that
the situation here is quite opposite to the reweighting at $\nu=0$:
one needs large ensembles exponentially in the size $L$,
to get signals at small $\mu$, while the simulations at large $\mu$
becomes easier.

We also show in Fig.~\ref{fig:Phase}
the reweighting factor for $L=16$ but with $\beta=2$,
intending one step toward the continuum limit
from $L=8$ and $\beta=1$. In this change of
parameters, the reweighting factors take the similar values to those
in the case with $L=8$ and $\beta=1$.

\begin{figure}[tb]
\begin{center}
\includegraphics[width=0.45\textwidth]{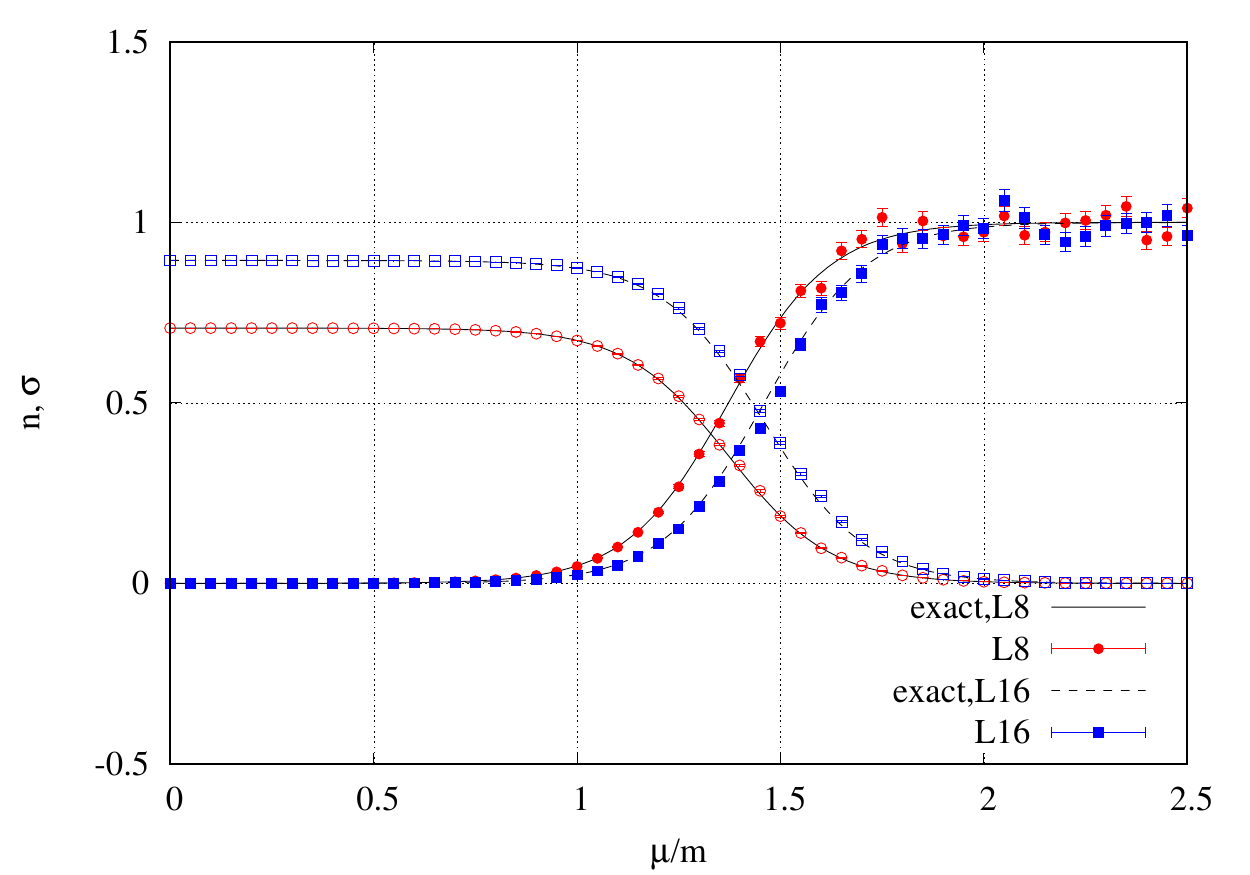}
\hfil
\includegraphics[width=0.45\textwidth]{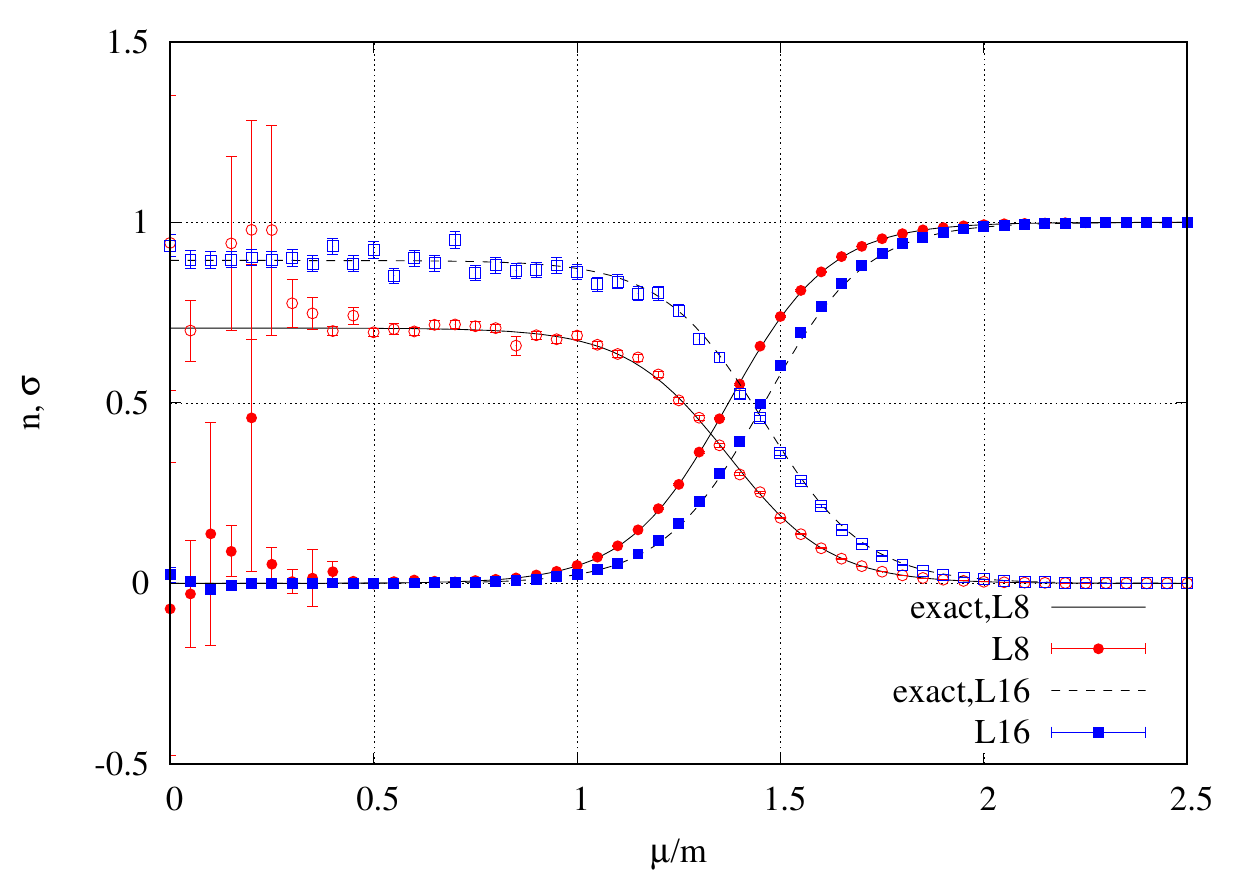}
\end{center}
\vspace{5mm}
\caption{
  Fermion number $n$ and scalar $\sigma$ densities as a function of $\mu/m$,
  evaluated with the reweighting method at $\nu=0$ (left)
  and $\nu/m=2.5$ ($3$) for $L=8, \beta=1, m=1$
  ($16, 2, 1/2$)
  in uniform-field model
  ($10^8/10^2$ samples collected with
  $\varepsilon=10^{-3}, \varepsilon_0=10^{-1}$).
}
\label{fig:RW-L8}
\end{figure}

\begin{figure}[tb]
\begin{center}
\includegraphics[width=0.45\textwidth]{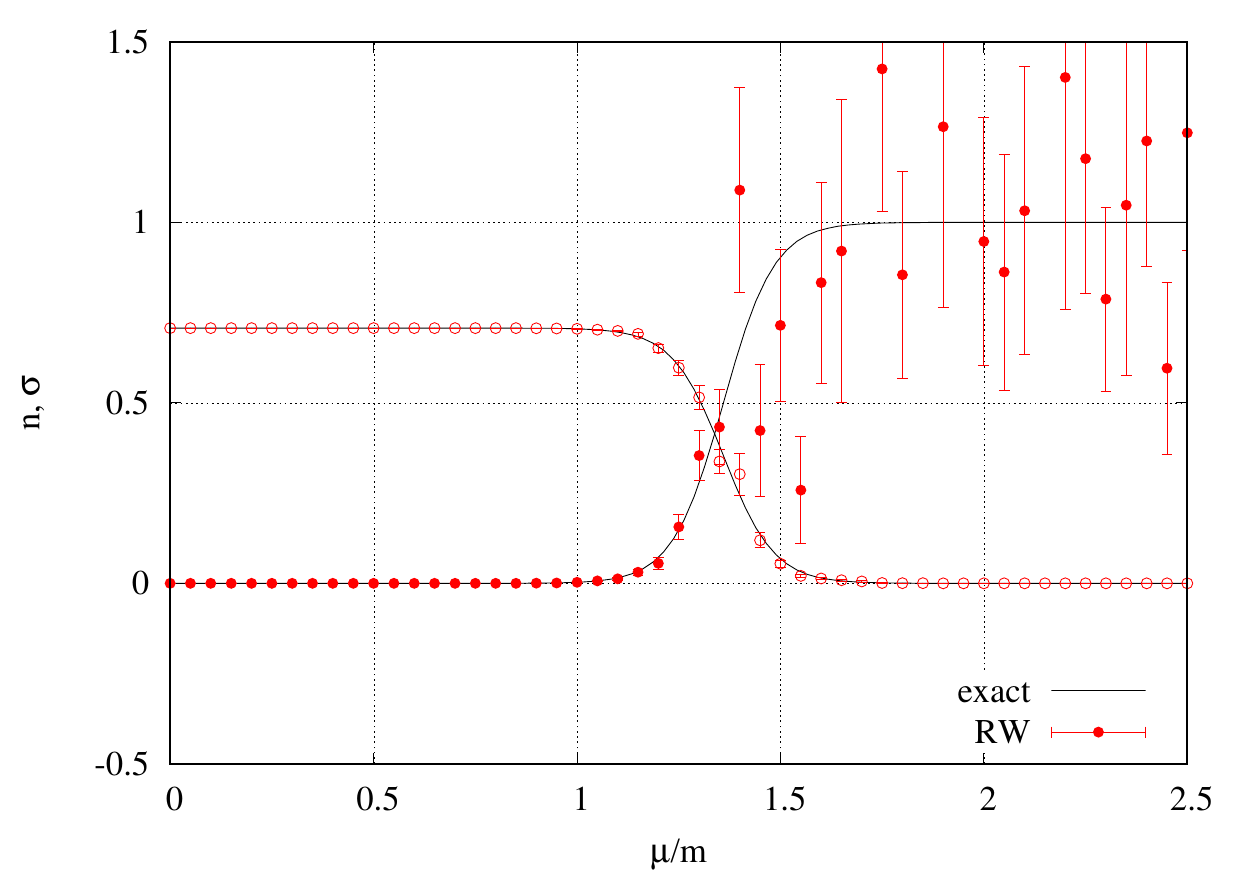}
  \hfil
  \includegraphics[width=0.45\textwidth]{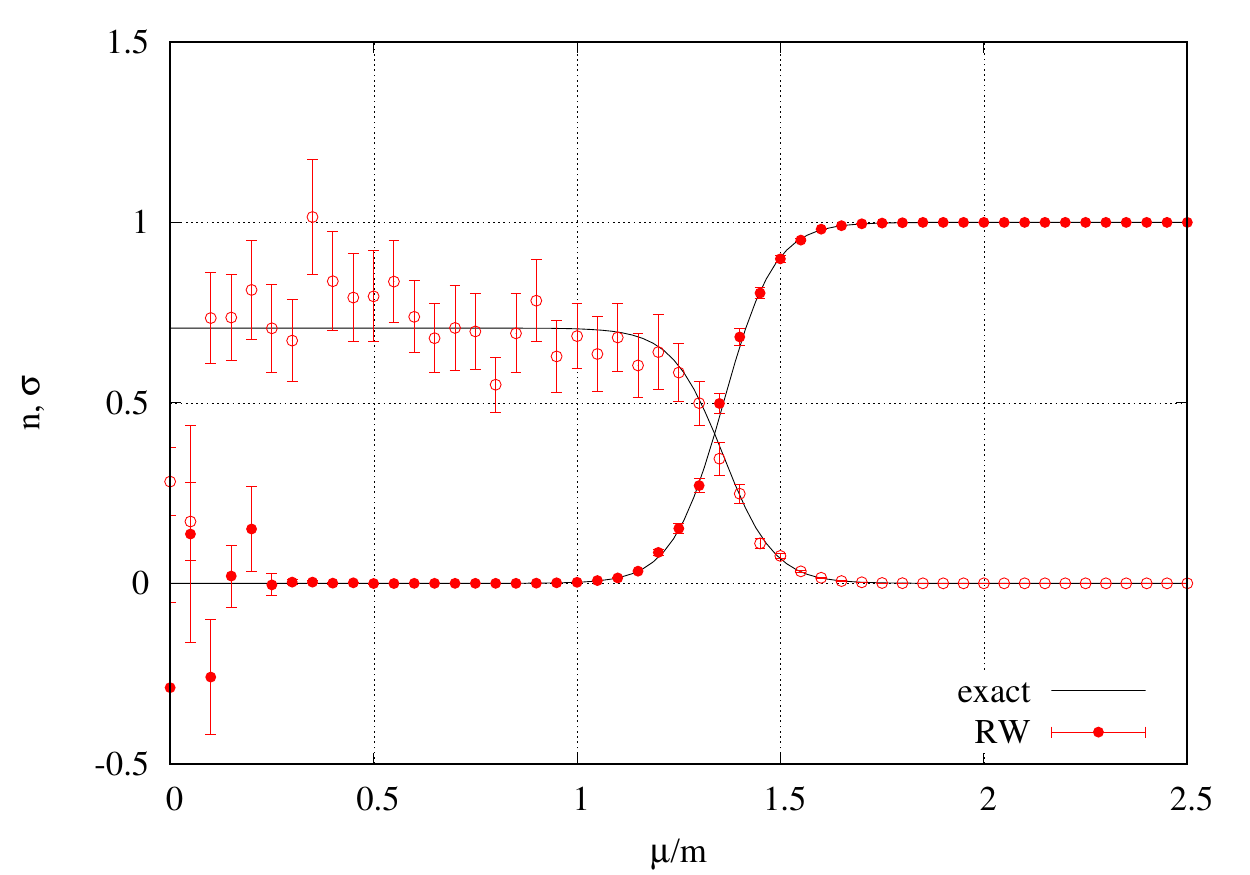}
\end{center}
\vspace{5mm}
\caption{
  Fermion number $\langle n\rangle$ and
  scalar $\langle \sigma \rangle$ densities as a function of $\mu/m$,
    evaluated with the reweighting method at $\nu=0$ (left)
    and $2.5$ (right)  for $L=16$, $\beta=1$, $m=1$
    in uniform-field model
    ($10^9/10^2$ samples are collected with
    $\varepsilon=10^{-3}$, $\varepsilon_0=10^{-1}$).
}
\label{fig:RW-L16}
\end{figure}

\subsection{Uniform-field model}

Now we apply the reweighting method to
the uniform-field model
with the parameters
$(L,\beta, m)=(8,1,1)$ 
with choosing the reference chemical potential
to $\nu/m=0$ and $2.5$.
As seen in Fig.~\ref{fig:CL-results},
the value $\mu/m=2.5$ is well above the crossover
region, where CL method gives the correct results.

For this parameter set, the phase average
$\langle \rme^{\rmi \phi} \rangle_{\rm p.q.}$ or
$\langle \rme^{\rmi \varphi} \rangle_{\nu \rm p.q.}$
becomes as small as of order $O(10^{-2})$, which implies
that we will lose numerical accuracy by this factor in the reweighting.
Thus, in order to detect significant signals,
we collected $10^6$ samples 
in the CL simulations of $10^8$ steps 
with the step size $\varepsilon=10^{-3}$
by making measurements every 100 interval steps.

In Fig.~\ref{fig:RW-L8},
results of reweighting at $\nu=0$ (left panel) and
$2.5$ (right panel) for $L=8$ are presented with red circles.
The bars indicate the statistical uncertainties assuming that
the samples with interval $\Delta \tau = 100 \varepsilon=0.1$
are independent.
We find that the reweighting method reproduces the correct results
reasonably for the fermion number $\langle n\rangle$
and scalar $\langle \sigma \rangle$ densities.
If we look closer in the crossover region $\mu/m = 1 \sim 2$,
we find that
the reweighting at $\nu=2.5$ works slightly better than
the reweighting at $\nu=0$.
This can be understood from the smallness of the phase
factors, $\langle \rme^{\rmi \phi}\rangle_{\rm p.q.}$ and
$\langle \rme^{\rmi \varphi}\rangle_{\nu,{\rm p.q.}}$,
presented in Fig.~\ref{fig:Phase}.

The results of the same model on the finer lattice ($L=16$, $\beta=2, m=1/2$)
at the one step toward the continuum limit,
are also shown with squares in Fig.~\ref{fig:RW-L8}.
We recognize here that the reweighting method works with similar 
quality between $L=8$ and $16$.

We perform the same simulations for lower-temperature case,
$L=16, \beta=1, m=1$, and
show the result in Fig.~\ref{fig:RW-L16}.
The statistical uncertainties become substantially larger
in this case than in the previous case.
This is again consistent with the smallness of the phase average
shown in Fig.~\ref{fig:Phase}. The ensemble size $10^{6}$ is
still insufficient to assure a good accuracy of the numerical
results. The size required for obtaining the significant numerical accuracy
will increase exponentially with $L$.
We notice here that
in the crossover region of $\mu$
the reweighting results at $\nu=2.5$ have smaller uncertainty bars 
than those at $\nu=0$ in Fig.~\ref{fig:RW-L16}.

\subsection{Thirring model in (0+1) dimensions}

\begin{figure}[tb]
  \begin{center}
    \includegraphics[width=0.45\textwidth]{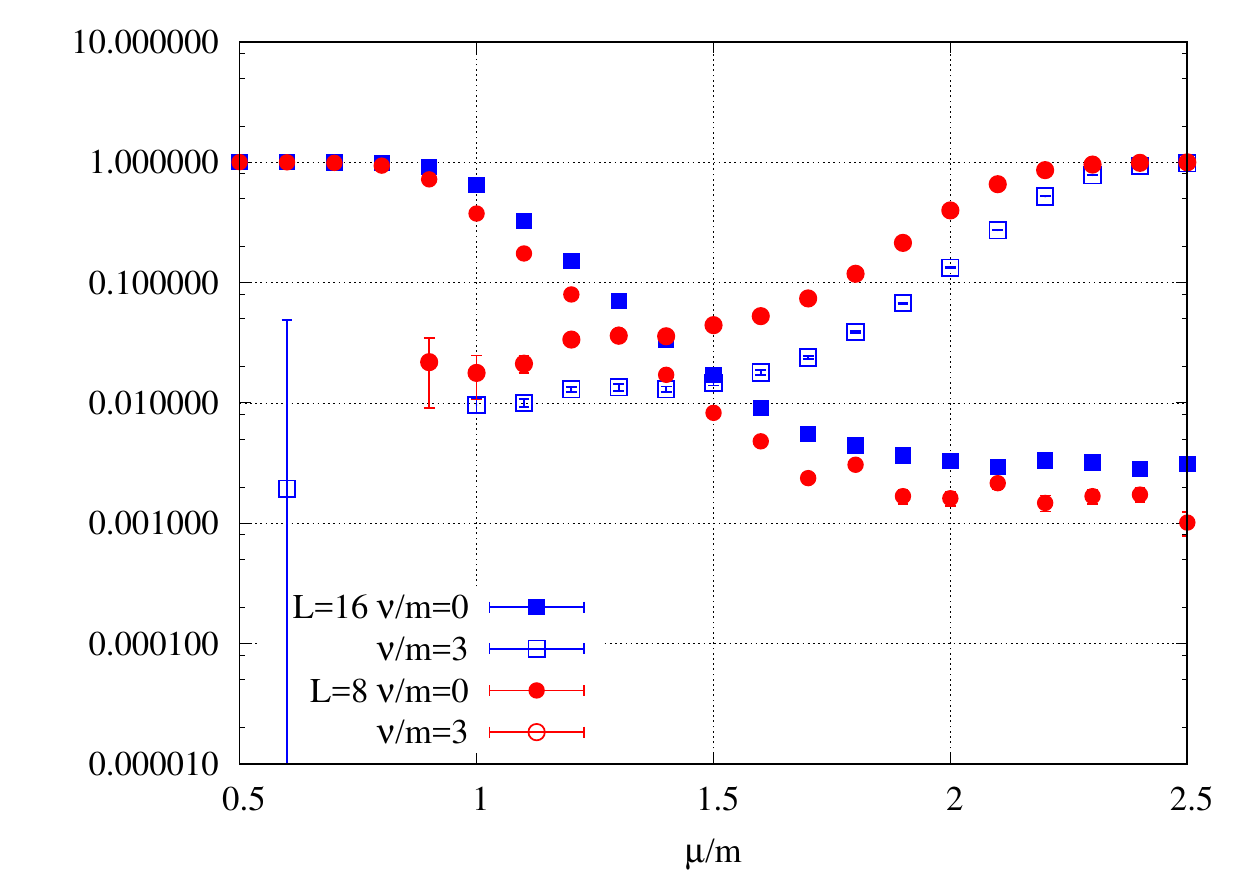}
    \hfill
    \includegraphics[width=0.45\textwidth]{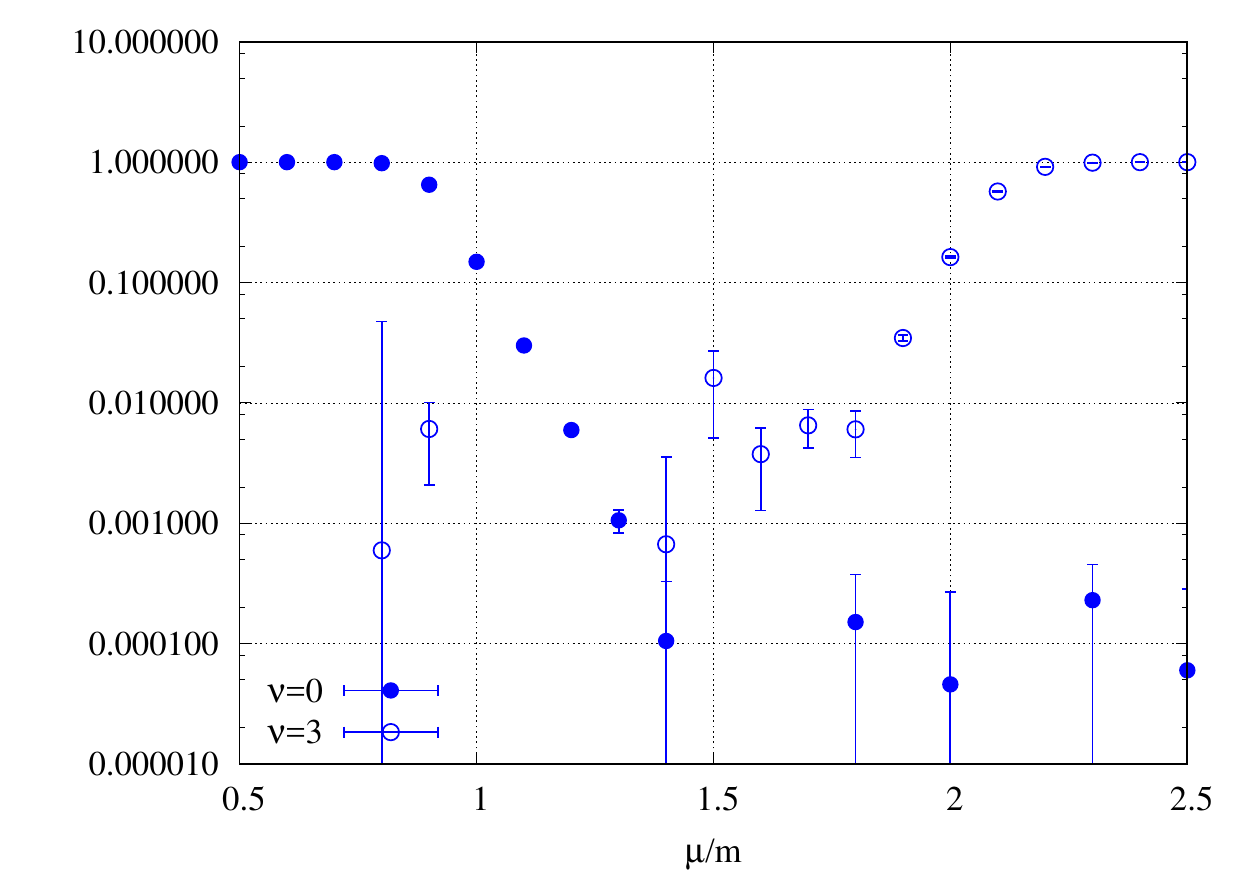}
  \end{center}
  \vspace{5mm}
  \caption{
    Expectation values of 
    $\langle \rme^{i \varphi} \rangle_{\nu\rm p.q.}$ as a function $\mu/m$
    in 0+1 dimensional Thirring model, evaluated in Langevin simulations.
    Left: $(L,\beta,m)=(8, 1, 1)$ and $(16,2, 1/2)$.
    Right: $(16, 1, 1)$.
}
\label{fig:PhaseFactor-0+1}
\end{figure}

We collected  $10^7$ samples with 200 interval steps
($\varepsilon=5\cdot 10^{-4}$,
 $\varepsilon_0=5\cdot 10^{-2}$)
to draw the figures in this subsection.
First we show in Fig.~\ref{fig:PhaseFactor-0+1}
the expectation value of the phase average,
the CL counterpart of Eq.~(\ref{eq:PhaseFactor2}).
The left panel shows the results for $L=8$ and $\beta=1, m=1$.
We see that 
the phase average obtained in the phase quench simulation
at $\nu=0$ (filled circles)
decreases to about $10^{-3}$ at large $\mu$. 
In contrast, 
the phase average evaluated with the reference parameter
$\nu=3.0$ (open circles)
becomes a few times $10^{-2}$ in the crossover region
but even turns to be negative with substantial errors
for $\mu \lesssim 1$.
On the finer lattice, $(L,\beta,m)=(16, 2,1/2)$,
the result for the phase average
evaluated with $\nu=0$ ensemble (filled squares) becomes
a bit larger in large $\mu$ region,
while the phase factor evaluated with $\nu/m=3.0$ ensemble (open squares)
takes smaller values in the crossover region,
as compared to $L=8$ case.
This results suggest that the sample points more than $10^6$
will be required to study these system with the CL simulations,
if the statistical uncertainty scales with square root of the sample size.

However, when we decrease the temperature by changing
$L=16$ from $L=8$ with $\beta=1, m=1$ fixed,
the phase average $\langle \rme^{\rmi \varphi} \rangle$
becomes smaller by one order of magnitude
with substantially larger statistical errors,
as is shown in the right panel of
Fig.~\ref{fig:PhaseFactor-0+1}.
We performed CL simulations with $\nu=0$ and 3
to generate the reference ensembles of size $10^{7}$,
but after reweighting we did not obtain any significant results
in the crossover region.
This result suggests that 
we need to search
more efficient reference parameters $\beta$ and $m$ in addition to $\mu$,
in order to apply the reweighting method at lower temperature,
which we leave for future study.

\begin{figure}[tb]
  \begin{center}
    \includegraphics[width=0.45\textwidth]{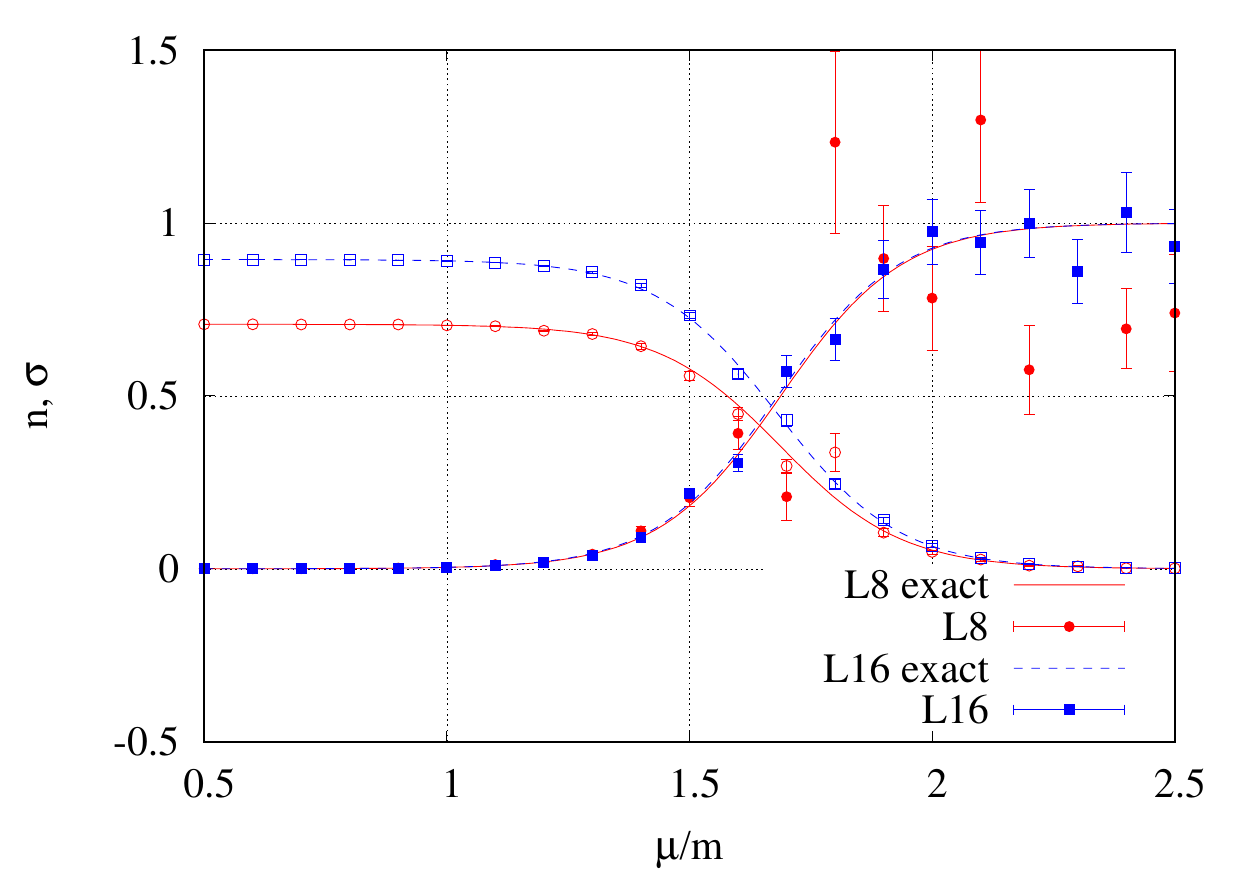}
    \hfill
    \includegraphics[width=0.45\textwidth]{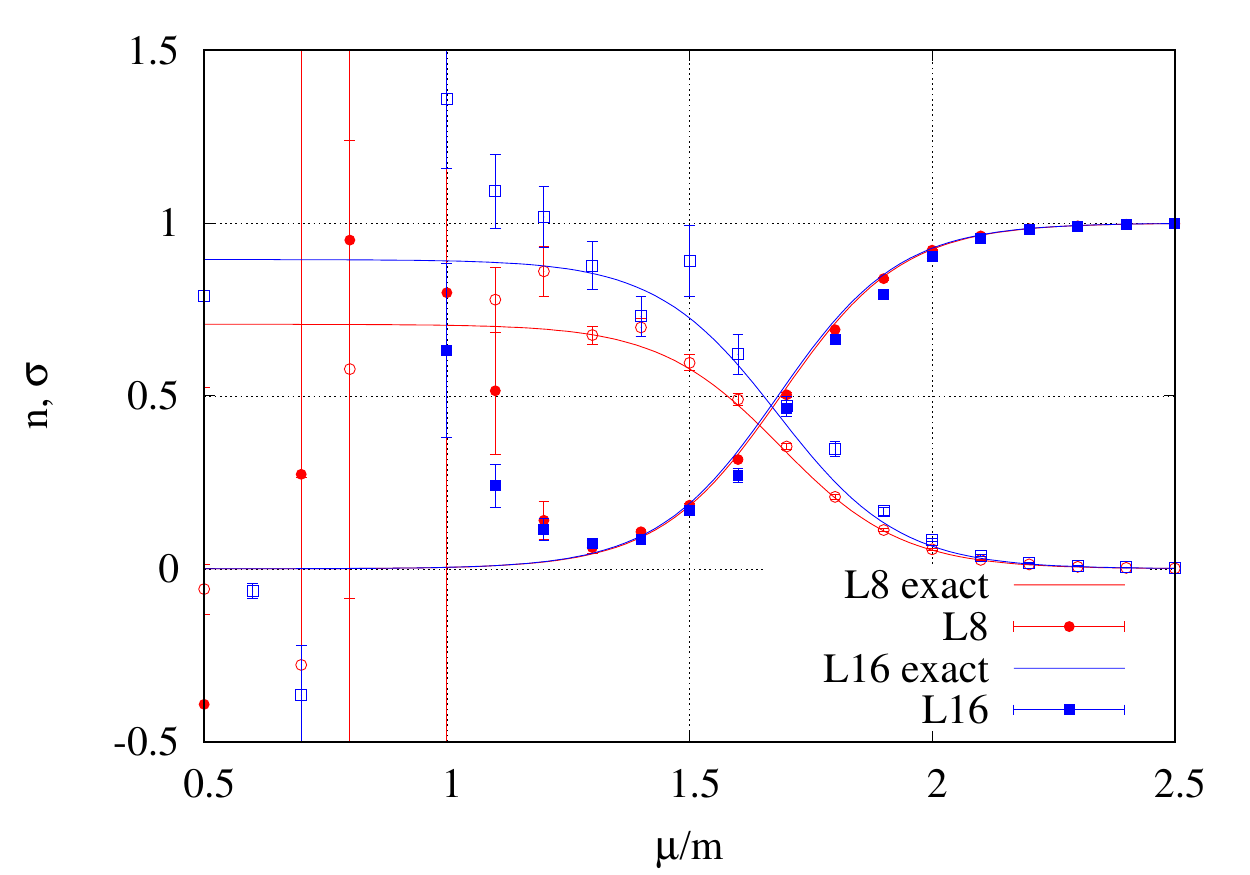}
  \end{center}
\vspace{5mm}
\caption{Number $\langle n\rangle $ and
  scalar $\langle \sigma\rangle $ densities obtained
    by reweighting at $\nu/m=0$ (left) and $3$ (right)
    for 0+1 dim Thirring model
    with $(L,\beta,m)=(8,1,1)$ (red circles)
    and $(16, 2, 1/2$) (blue squares).
}
\label{fig:RW-full-L8b1}
\end{figure}

\begin{figure}[tb]
  \begin{center}
    \includegraphics[width=0.45\textwidth]{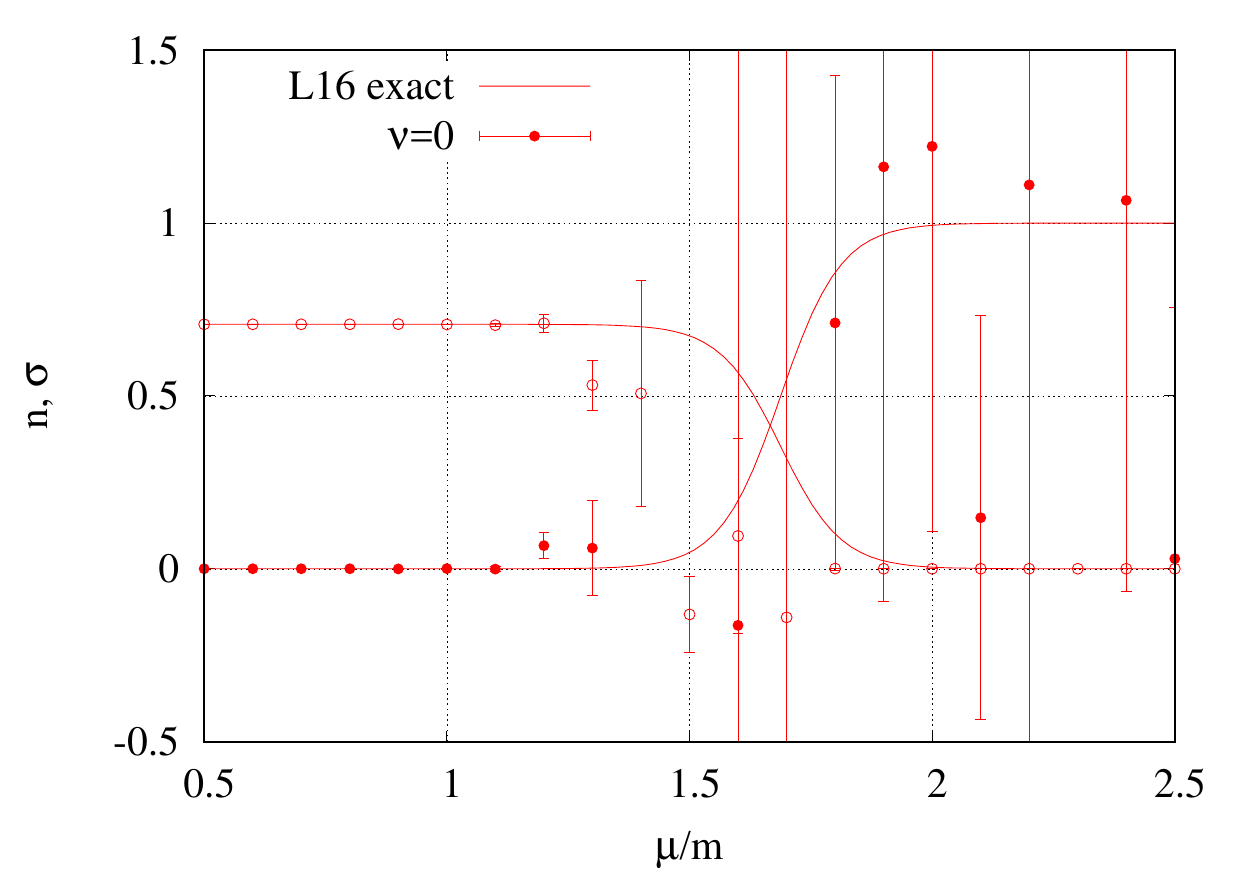}
    \hfill
    \includegraphics[width=0.45\textwidth]{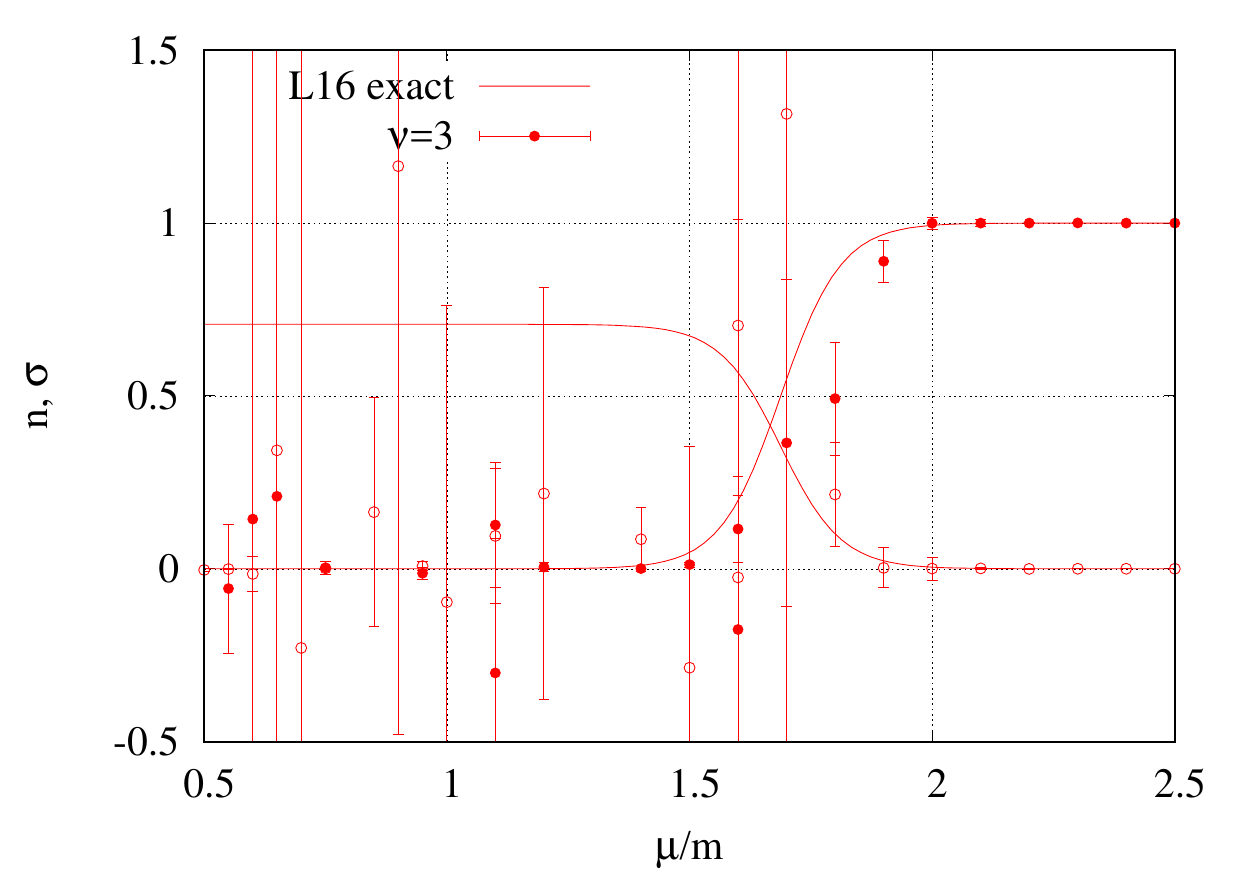}
  \end{center}
  \vspace{5mm}
\caption{Reweighting result of 0+1 dim Thirring model
  with $\beta=1$, $m=1$, $L=16$.
  (data points to be added)}
\label{fig:RW-full-L16b3}
\end{figure}

\section{Deformation method}

  Tsutsui and Doi\cite{Tsutsui:2015tua}
  proposed a model deformation for a simple fermionic model
  to avoid the problem of the determinant zeros,
  by noting an empirical relation between
  the multi-thimble structure of the model and
  the failure of the CL simulation.
  In this section, we examine the deformation method by applying
it to the uniform field model.

\subsection{Exact expressions}

  In Ref.~\cite{Tsutsui:2015tua}
  an extra term is added to the polynomial term of the original model
  so that the deformed model has
  a single thimble covering the integration contour
  and the CL mothod becomes applicable to the model.
  Then the observables of the original model
  are expressed
  exactly in terms of those of the deformed models,
  by utilizing the following identity
  \footnote{The second expression is due to S.~Shimasaki.}:
\begin{align}
\label{eq:TDidentity}
\text{(I): } \qquad
   \left < O \right >_f = &    \left < O \right >_{f+g} +
\left  (\left < O \right >_{f+g} - \left < O \right > _g  \right )
   \frac{\left < g  \right > _0 } {\left <  f \right >_0}
  \notag \\
  = & \left < O \right >_{f+g} +
  \left (\left < O \right >_{f+g} - \left < O \right > _g \right )
  \left < \frac{ f }{  g } \right >_g^{-1}
\, ,
\end{align}
where $\langle O \rangle_{f, f+g}$ denote respectively
the expectation values taken with the partition functions,
\begin{align}
Z_f  \equiv \int \frac{dx}{2\pi}  \rme^{-S_b(x)}  f(x)
\, ,
\qquad\,
Z_{f+g}  \equiv \int \frac{dx}{2\pi}  \rme^{-S_b(x)}  \big [ f(x)+g(x) \big ] 
\end{align}
with certain functions, $f(x)$ and $g(x)$.
In the present model, we identify $f(x)=\det D_0(x; \mu)$, and
among many possibilities for $g(x)$ we choose here
$g(x)=\det D_0(x; \mu_g)$ with $\mu_g$ a reference chemical potential.
(Hereafter, we denote the physical chemical potential $\mu=\mu_f$.
The expectation value $\langle O \rangle _0$ is taken with
the fermion quenched model without $f$ and $g$ factor.
This identity will be beneficial if one can find $g(x)$
such that CL simulation produces the correct results for $Z_{g, f+g}$
even when it does not work for $Z_f$.
Vanishing $\left < g \right>_0=0$ is the most beneficial case. 
But there is no general proof for the existence of such a special choice
of $g(x)$ for a given model.

\begin{figure}[t]
\begin{center}
\includegraphics[width=0.35\textwidth]{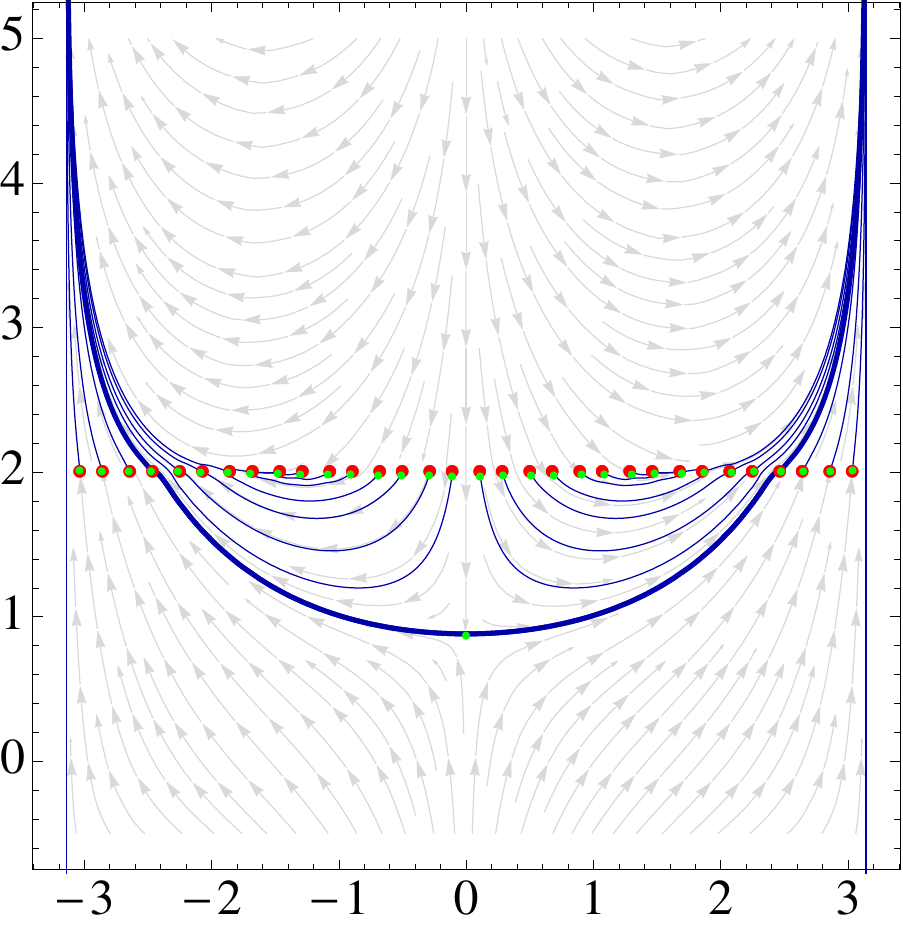}
\hfil
\includegraphics[width=0.35\textwidth]{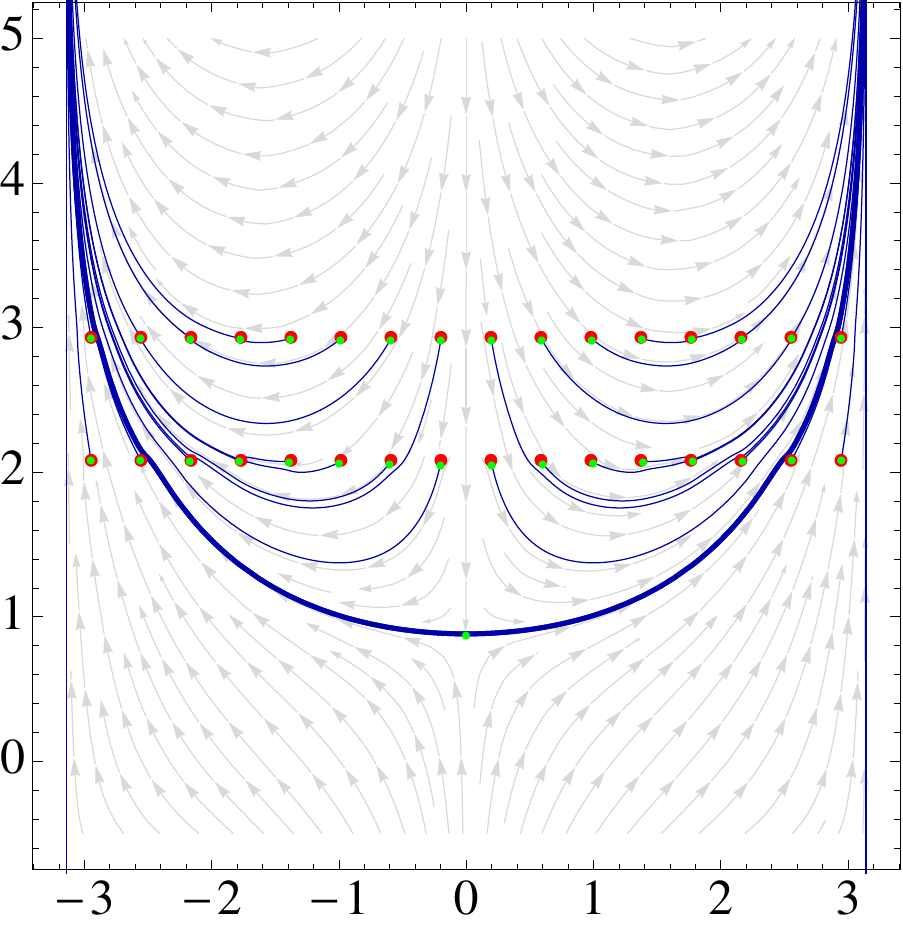}
\end{center}
\caption{
Thimbles (blue lines) and gradient flow fields in the $z=(x,y)$ plane for the modified model $Z_{f+g}$ for $\mu_f=1$ (left) and 2 (right)
($\mu_g=3$; $\beta=1, m=1,L=16$). The zeros and critical  points are shown in red and green, respectively.
}
\label{fig:DT-Thimble}
\end{figure}

In the uniform-field model, 
we have seen that CL method gives the correct results for a large $\mu_f$,
where the thimble structure becomes simple and
the original integration contour is practically
covered with a single thimble.
By choosing the reference chemical potential $\mu_g$ large,
we can make the thimble structure
of the deformed model $Z_{f+g}$ (as well as $Z_g$)
simple for $\mu_f$ within the range considered here.
For $\mu_g=3$ we draw
the thimble structure of the deformed model $Z_{g+f}$ with
$L=16, \beta=1, m=1$ and $\mu_f=1$ (left) and 2 (right)
in Fig.~\ref{fig:DT-Thimble}.
All the zero points locate well above the real axis
in the imaginary direction and the single dominant thimble
practically covers the integration domain for $\mu$ in the range
considered here
(see Appendix A for analytic expressions for the zero points).

The sign problem in this expression remains in the multiplier
$\left <g\right>_0/\left < f\right>_0$
appearing in Eq.~(\ref{eq:TDidentity}):
Both the integrands in the denominator and numerator
become very oscillatory for large $\mu_{f,g}$.
This multiplier is actually 
the inverse of the reweighting factor
as is indicated in the second line of Eq.~(\ref{eq:TDidentity}).

We evaluate the exact expression (\ref{eq:TDidentity}) with CL method
for the observables $n(x;\mu)$ (\ref{eq:n0})
and $\sigma(x;\mu)$ (\ref{eq:sigma0}).
We notice here, however, that CL simulations
may hit the poles of these observables
depending on the value of $\mu_f$, although
the poles of the drift term of
the deformed model are avoided by choosing $\mu_g$ large.
The CL simulation will not be justified
for such values of $\mu_f$ in this approach.

Concerning this point,
we like to remark 
that if we adopt also for the number density observable
the deformed expressions,
\begin{align}
  \label{eq:tilden_fg}
  \tilde n_{f+g} (x) & \equiv \frac{\sinh L(\mu_f + \rmi x )+\sinh L(\mu_g+\rmi x)}
         {\det D_0(x;\mu_f)+ \det D_0(x;\mu_g)}
\; ,
\\
\label{eq:tilden_g}
\tilde n_{g}  (x)  & \equiv  \frac{\sinh L(\mu_g+\rmi x)} {\det D_0(x;\mu_g)}
\, ,
\end{align}
then we find another, but similar identity:%
\begin{align}
\label{eq:Kidentity}
\text{(II): } \qquad
   \langle  \tilde  n_f \rangle _f = 
      \langle  \tilde n_{f+g} \rangle _{f+g} + 
    \Big ( \langle  \tilde n_{f+g} \rangle _{f+g}  -   \langle  \tilde n_g \rangle _g \Big ) 
\left \langle     \frac{ \det D_0(\mu_f)  } { \det D_0(\mu_g) } \right >_g^{-1}
        \; .
\end{align}
We also have a corresponding identity for $\left <\sigma\right >$
with deformed expressions $\tilde \sigma_{g,f+g}(z;\mu)$.
The expression (II) has two good properties;
(i) the pole locations of $\tilde n_{g, f+g}$ are the same as those of
the drift term, which are avoided in the CL evolution
by taking $\mu_g$ large enough,
and 
(ii) we have analytic expressions for $\left <\tilde n_{g,f+g} \right>$
to be compared with the numerical results.

\subsection{Numerical results}

We present numerical results of the deformation method applied to
the uniform-field model for $L=8$ and $16$ with $\beta=1, m=1$.
In the simulations
we chose here
$\varepsilon=10^{-4}$ and
$\varepsilon_0=10^{-2}$,
and collected the samples $N=10^6$ with $10^2$ interval steps.

\subsubsection{Case (I)}

\begin{figure}[t]
\begin{center}
\includegraphics[width=0.325\textwidth]{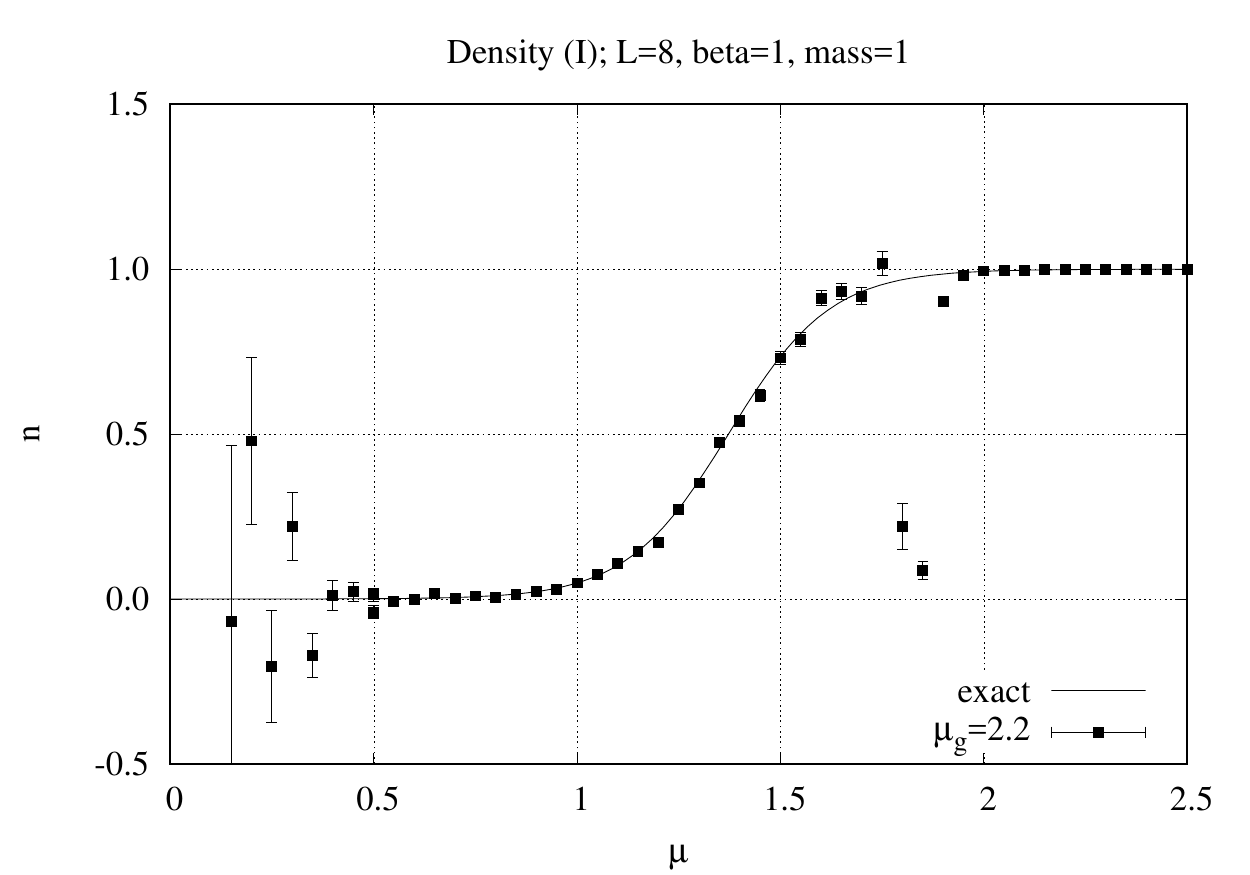}
\hfil
\includegraphics[width=0.325\textwidth]{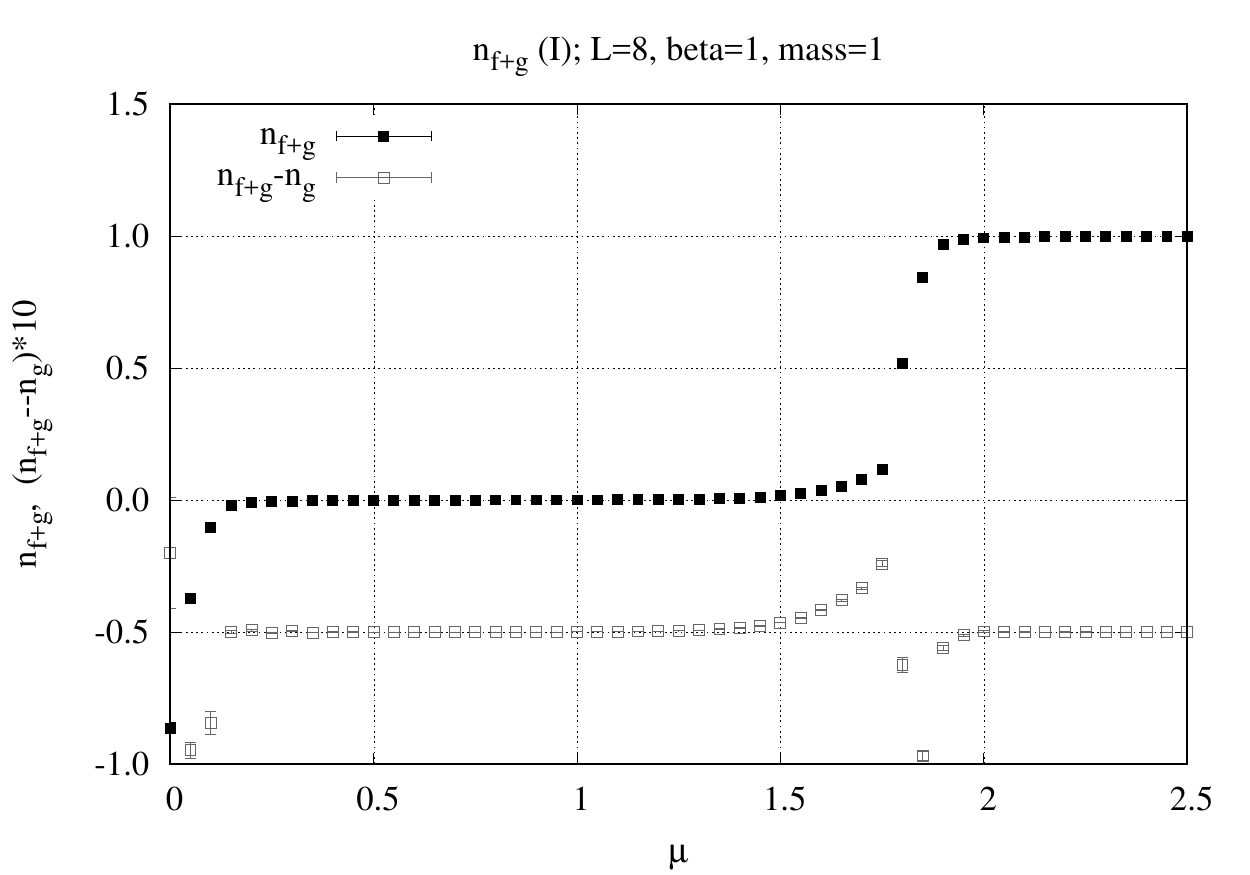}
\hfil
\includegraphics[width=0.325\textwidth]{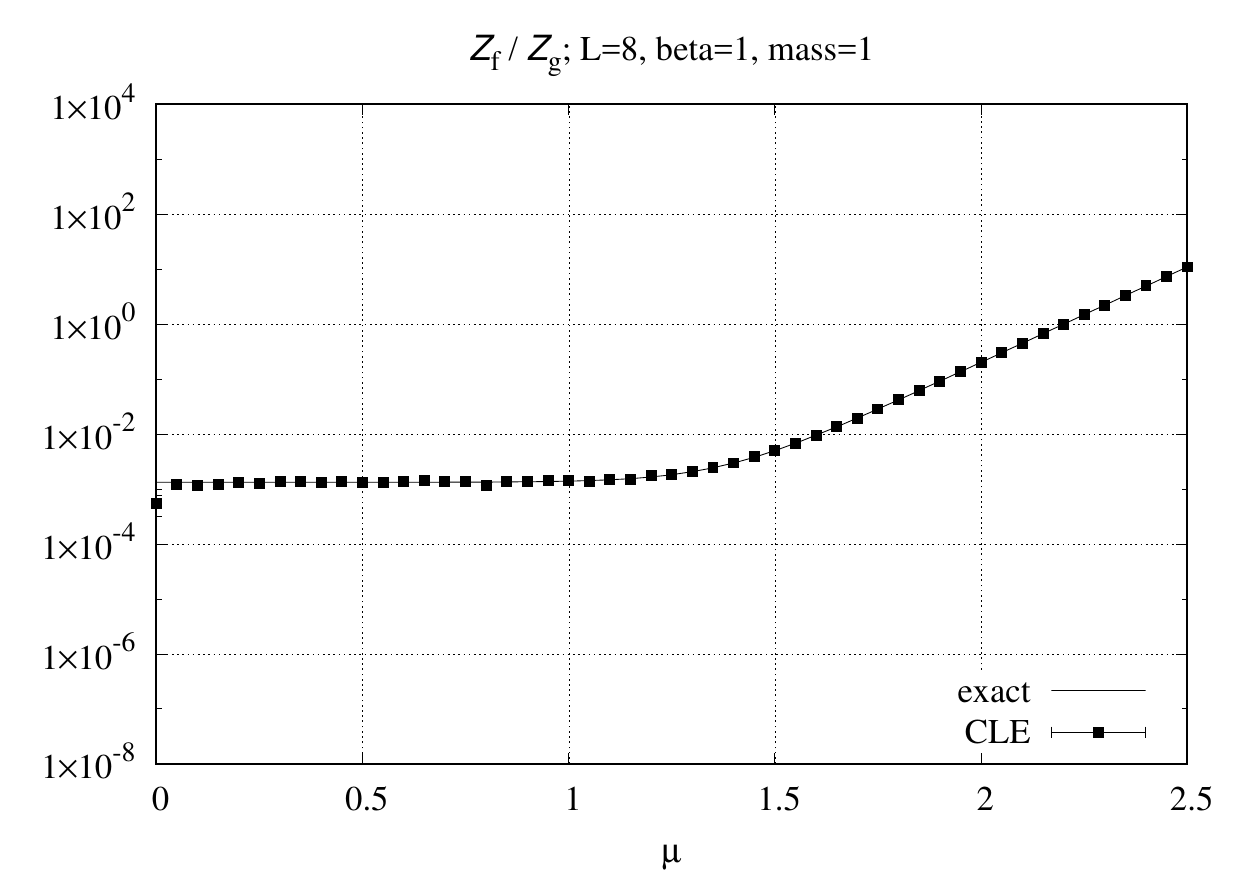}

\includegraphics[width=0.325\textwidth]{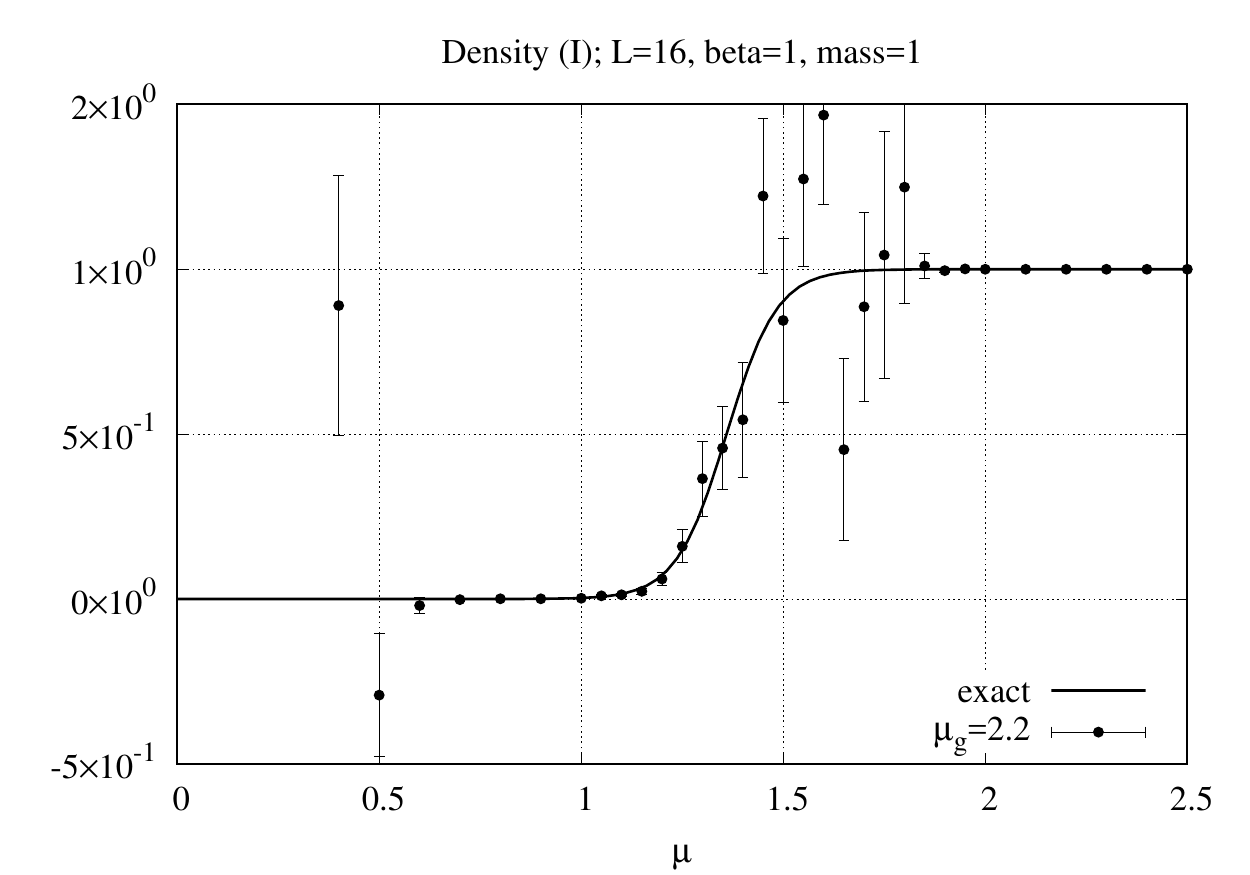}
\hfil
\includegraphics[width=0.325\textwidth]{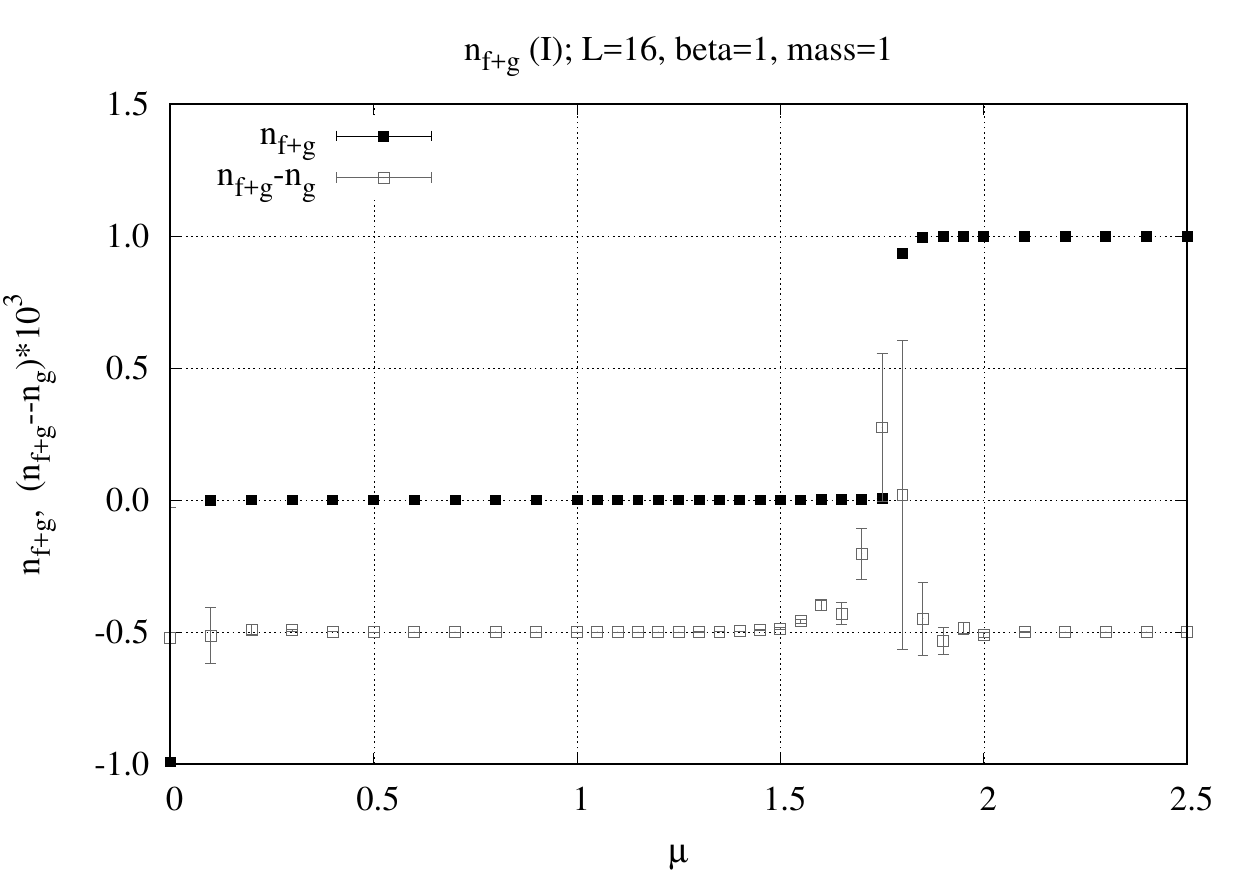}
\hfil
\includegraphics[width=0.325\textwidth]{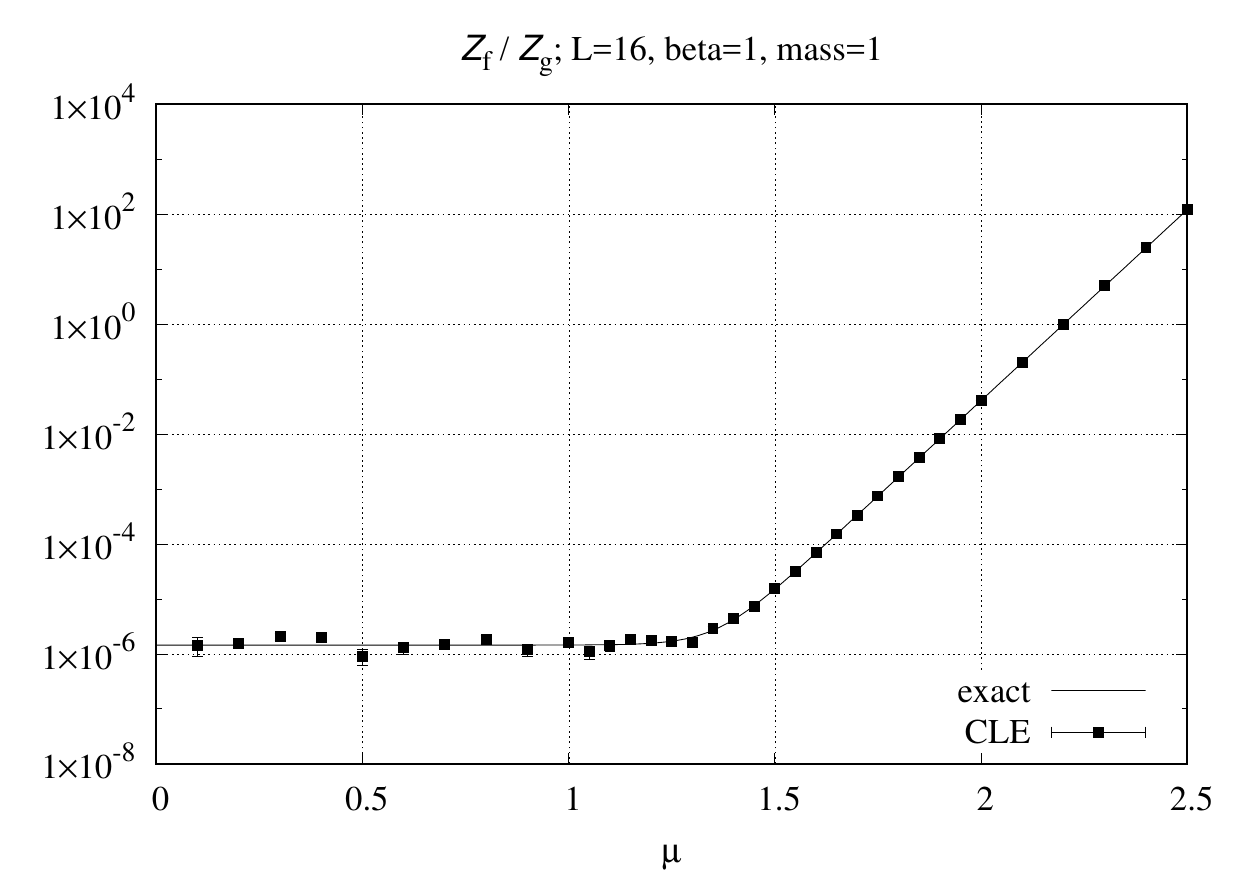}

\end{center}
\caption{
  Fermion number density obtained in CL simulation using the formula (I) [left], its decomposition $\langle n \rangle_{f+g}$ and
$\langle n \rangle _{f+g} - \langle n \rangle_g$ (offset by $-0.5$ for visibility) [middle], and reweighting factor [right]
for $L=8$ (upper) and 16 (lower) with $\beta=1$, $m=1$, and $\mu_g=2.2$.
}
\label{fig:DT-density}
\end{figure}

We display the numerical results of the deformation method (I)
applied to the uniform field model
for $L=8$ in the upper row of Fig.~\ref{fig:DT-density},
and those for $L=16$ in the lower row. We choose here $\mu_g=2.2$.
For this value of $\mu_g$ the integration contours  practically get
covered by the dominant thimbles ${\cal J}_{z_0}$ of $Z_{g,f+g}$, respectively.
The left panels show the density obtained $\left < n \right >_{f}$
by formula (I),
and the middle and right panels show the intermediate quantities appearing on
RHS of formula (I).

We show $\left < n \right >_{f+g}$ in middle panels of Fig.~\ref{fig:DT-density}
as a function of $\mu_f$. 
The CL results for both $\left < n \right >_{f+g}$ and $\left <n\right >_g$
actually show abrupt changes around $\mu_f \sim 1.8$ (and $\sim 0$).
We have confirmed that they
are almost consistent with the values obtained
by integration along the thimbles of $Z_{g,f+g}$.
In the thimble integration, these densities make jumps as
the poles in the integrand (\ref{eq:n0}) cross the integration contours,
but they show abrupt but smooth change here because the CL ensembles extend
over the complex plane, not on a line.
We also show with an offset $-0.5$
the difference $\left < n \right >_{f+g} -\left <n\right >_g$ there,
which has a spike because the abrupt change of each term occurs
at different $\mu_f$ because the CL ensembles for $Z_{g, g+f}$ are not identical
\footnote{
  If we evaluate $\left < n \right >_{f+g}$ and $\left <n\right >_g$ by integration along the thimble,
  the poles of their integrands cross the thimbles at different values of $\mu_f$
  because the thimbles of $Z_{g, f+g}$ are slightly different.
  As a result, the subtracted density $\left < n \right >_{f+g} -\left <n\right >_g$ shows a spike.
  In order for (I) to hold,
  one must choose the same integration contour in each term on RHS
  so that the pole contributions must cancel out on LHS of Eq.~(\ref{eq:TDidentity}).
}.

As shown in left panel of Fig.~\ref{fig:DT-density},
the numerical result of (I) for $L=8$
reproduces the crossover behavior fairly well except
in the regions $\mu \sim 1.8$ and $\sim 0$, where we see the spike structure.
In the CL simulations,
a complete cancellation of the abrupt changes in the subtraction term on RHS
is impossible to reproduce the exact result.
Moreover the poles in the expression of the observable make the correctness
of CL simulation very dubious.

In formula (I), the gradual $\mu_f$ dependence of the exact curve
should be reproduced by a delicate combination of 
$\left <n\right >_{f+g} - \left <n\right >_g$ and
$(Z_f/Z_g)^{-1}=\left < D_0(\mu_f)/D_0(\mu_g)\right>_g^{-1}$,
because $\left < n \right >_{f+g}$ is almost flat
except for the abrupt changing point of $\mu_f$.
It is statistically challenging to evaluate accurately
the subtraction $\left <n\right >_{f+g} - \left <n\right >_g$
and the reweighting factor (right panel)
for larger size systems.
Note the magnification factors $10$ and $10^3$ in the plot
of the subtracted density for $L=8, 16$, respectively.
The numerical severity is seen clearly in the result for $L=16$ as shown in
the lower panels of Fig.~\ref{fig:DT-density}, where
we have much larger statistical errors.

\subsubsection{Case (II)}

\begin{figure}[t]
\begin{center}
\includegraphics[width=0.32\textwidth]{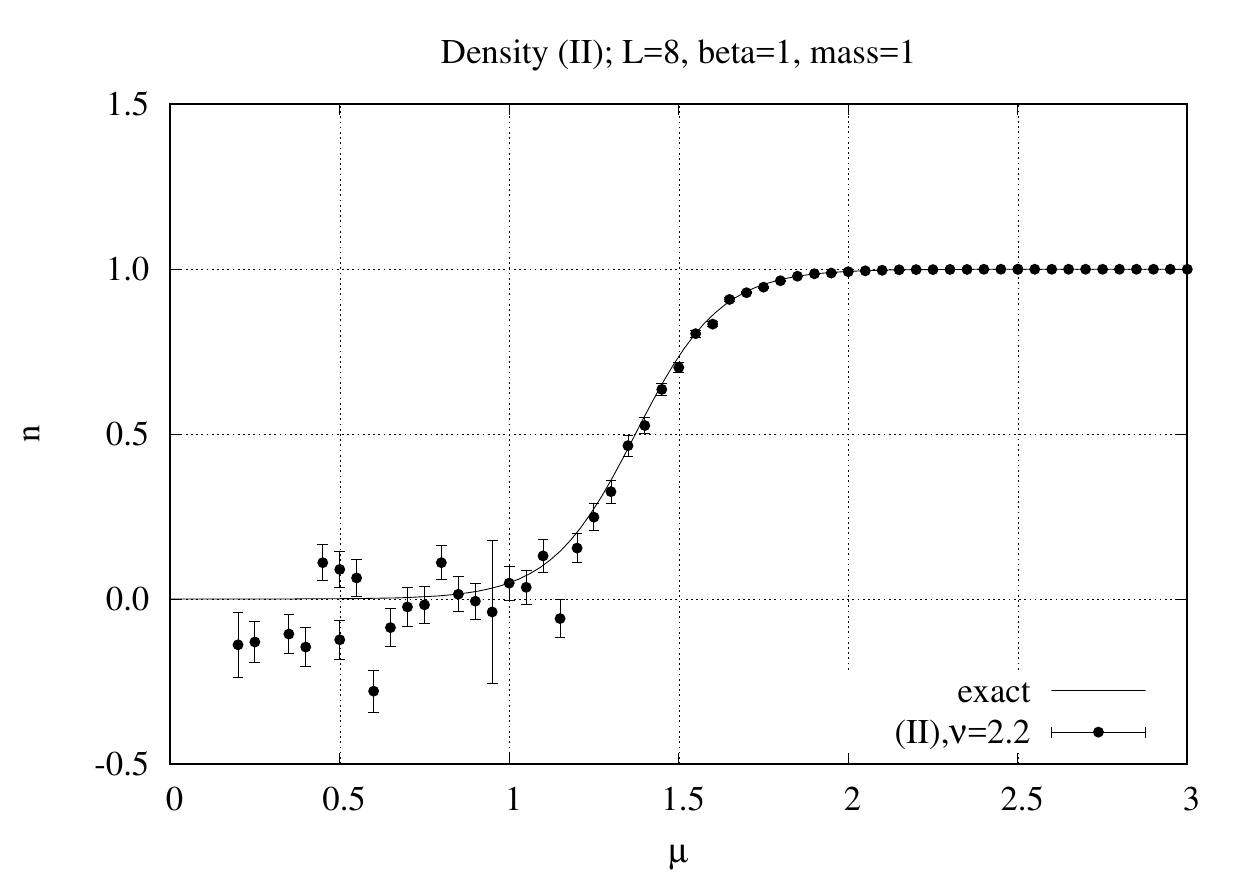}
\hfil
\includegraphics[width=0.32\textwidth]{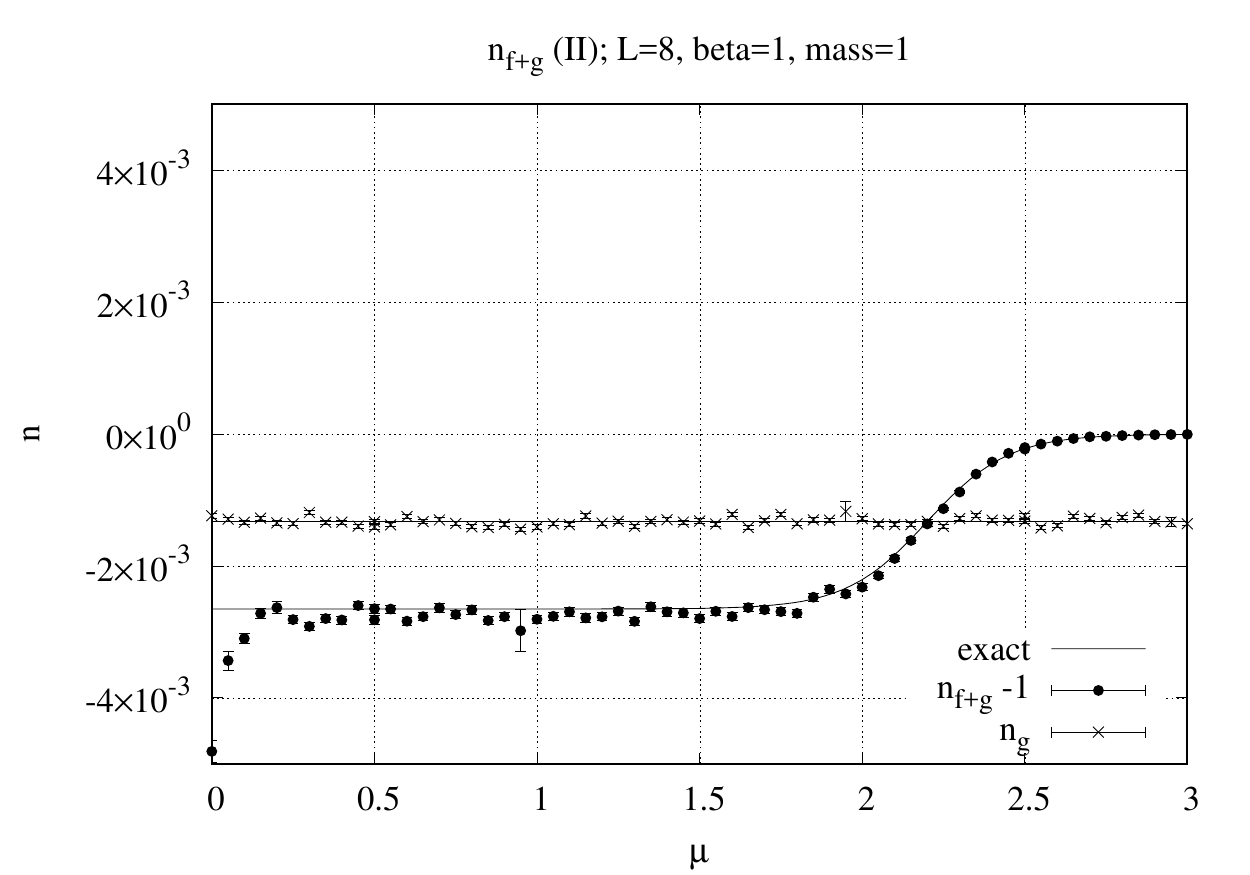}

\includegraphics[width=0.32\textwidth]{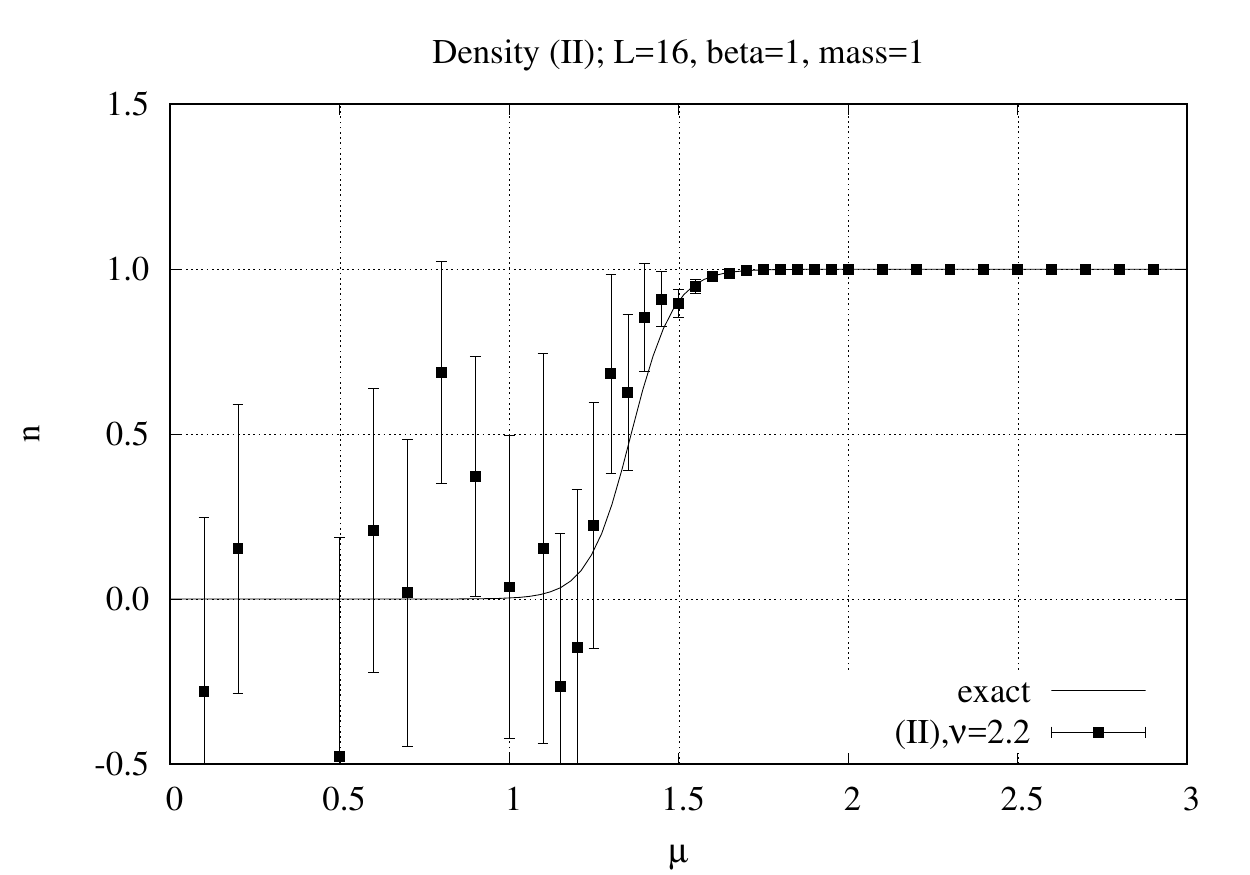}
\hfil
\includegraphics[width=0.32\textwidth]{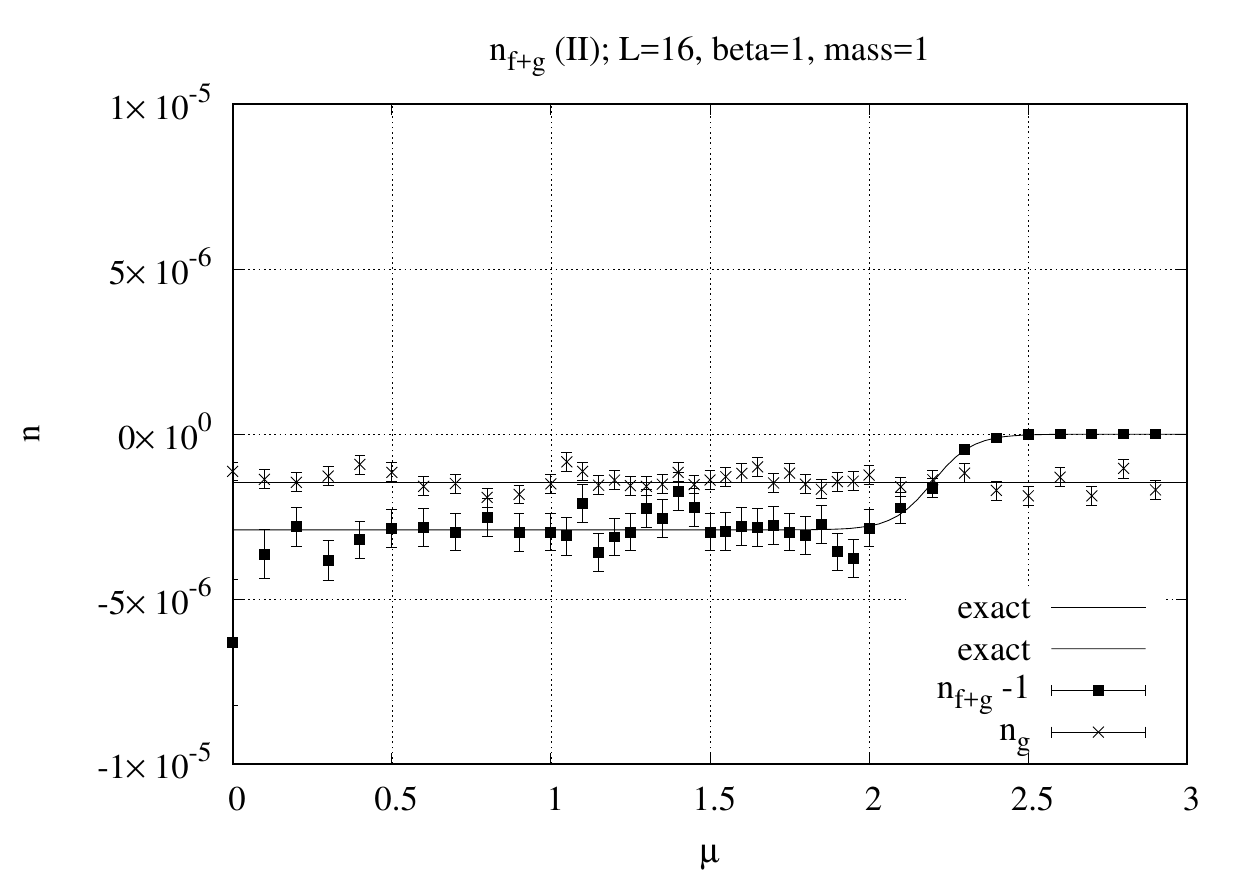}

\end{center}
\caption{
  Fermion number density obtained in CL simulation using formula (II) [left],
  its decomposition $\left < n_{f+g}\right > - 1$ and $\left < n_g \right >-1$ [right]
  for $L=8$ (upper) and 16 (lower) with $\beta=1, m=1$ and $\mu_g=2.2$.
}
\label{fig:KDT-density}
\end{figure}

In case (II), expressions for the observables $\tilde n_{g, f+g}$ have the
poles at the same locations as the drift term,
which are avoided in the CL simulation by taking $\mu_g$ large.
Thus the fermion number densities, $\langle \tilde n \rangle_{g, f+g}$,
will not show any abrupt dependence on $\mu_f$, unlike to case (I).
We show in Fig.~\ref{fig:KDT-density}  the fermion number density
$\langle \tilde n \rangle_{f}$ obtained for $L=8$ (upper) and 16 (lower)
with $\beta=1$ and $m=1$, 
by using formula (II) and compare it with
the exact result [left] and its decomposition [right],
$\langle \tilde n \rangle_{ f+g} -1 $ (filled circles) and
$\langle \tilde n \rangle_{ g} -1 $ (crosses).
In Fig.~\ref{fig:KDT-density},
$\mu_f$ dependence of numerical result for each term
is smooth and very consistent with the exact result within the errors.

However, for the reference value $\mu_g$ large above the crossover value, 
both $\langle \tilde n \rangle_{g, f+g}$ become
nearly unity regardless of the value
of $\mu_f$ within the range of our interest
(see Eqs.~(\ref{eq:tilden_fg}) and (\ref{eq:tilden_g})).
After taking subtraction,
$\langle \tilde n \rangle_{ f+g} -\langle \tilde n \rangle_{ g}$,
we are left with a tiny number as seen in the right panel
of Fig.~\ref{fig:KDT-density}.
In order to get the fermion number density $\left < n_f\right >$,
we multiply to this density difference
the reweighting factor
$Z_f/Zg=\left < D_0(\mu_f)/D_0(\mu_g)\right>_g^{-1}$,
whose evaluation becomes more difficult at the smaller $\mu$ and the larger $L$.
Although formula (II) gives a fair result in the crossover region for $L=8$,
the problem becomes already challenging for $L=16$, as is seen in
Fig.~\ref{fig:KDT-density}.

\section{Summary}

We have studied applicability of the LC simulation method
to the Thirring model in (0+1) dimensions at finite $\mu$,
and its uniform-field variant.

The CL simulations reproduce the exact solution at small and large $\mu$.
However, they deviate from the exact one in the crossover region,
where the CL ensemble distribution has an overlap with the
neighborhoods of the determinant zeros, where
the drift force becomes singular.
This observation is consistent with the argument for correctness of the CL simulation,
given in Refs.~\cite{Nagata:2016vkn}
(see also \cite{Aarts:2017vrv}).

We have found
that the ensemble distribution localizes around the thimbles
in the CL simulations for the Thirring model, in which the critical points
of the relevant thimbles behave as attractive points of the drift flow field
and the zeros of the determinant become the saddle points of the flow.
It seems inevitable in this model that the CL ensemble has an overlap
with the determinant zeros whenever
the corresponding thimble integration has contributions from
multi thimbles connected via these zeros.

In order to extend the applicability of
the CL simulation to the crossover region,
we have attempted two methods, both of which make the thimble structure of the
model simple to cover the integration path with a single thimble in view of the
thimble integration.
But we have noticed that both methods
need the evaluation of the reweighting factor.
With the direct reweighting method at large reference chemical potential $\nu$,
we can reproduce the correct crossover behavior of the observables
as a function of physical chemical potential $\mu$.
In this sense, one can think of the ensemble generated by CL method physical
when CL simulation is successful and can be used for the reweighting.
In the deformation method, we also reproduce the correct result but
there appears an unphysical structure when the observables have
the poles coming from the original determinant zeros.
But one can remove this unphysical structure by employing the deformed
expressions for the observables in accord with the model.

For a large system size, however, both methods encounter a difficulty
in evaluating the reweighting factor; the phase factor average amounts to
a small value which scales exponentially with the system size.
In the deformation method, we need to evaluate the subtraction
of the two expectation values of the similar magnitudes,
which gives an additional limitation on the method.

Here we only choose the chemical potential $\mu$ as a reference parameter,
but one can consider other reweighting by taking other model parameters,
such as the mass $m$ or temperature $T$, etc.,
as reference parameters for the reweighting.
We leave the examination of the multi-parameter reweighting method
for future work.

The eigenvalue distribution of the fermion operator
of the models here in the configuration space complexified by CL method
is also intriguing quantity to be studied, as was discussed
in Ref.~\cite{Pawlowski:2013pje,Splittorff:2014zca,Mollgaard:2014mga}.

\vspace{5mm}
\noindent
    {\bf Acknowledgement}
    S.K.\  is very grateful to S.~Shimasaki for useful discussions on this work.
    This work was supported in part by Grants-in-Aid of JSPS,
    \# 16K05313 and 16K05343.

\vspace{5mm}


\appendix

\section{Determinant zeros in uniform-field model}

We show the zero points of the uniform-field model and those for
the deformed model.

First, the analytic structure of the reference system $Z_g$ is the same as $Z_f$
with a simple replacement of  $ \mu_f \to \mu_g$;
the zeros in the uniform-field model, $D(z_{\rm zero}, \mu)=0$
are given as\cite{Fujii:2015bua}
\begin{align}
  L  z_{\rm zero} =  (2n+1)\pi +\rmi L (\mu \pm \hat m),
\end{align}
and there appear critical points associated to these zeros
as seen in Fig.~\ref{fig:ScattPlots}.

In order to understand the thimble structure of the modified model, $Z_{f+g}$,  let us first find
the zeros of $D(z_{\text{zero}};\mu_f) + D(z_{\text{zero}};\mu_g)=0$.
To this end, it is convenient to introduce $\tilde z$ defined by
$z=\tilde z+ \rmi (\mu_f+\mu_g)/2$, in terms of which
$\mu_f+\rmi z =  - \delta +\rmi \tilde z$ and
$\mu_g+\rmi z =   \delta +\rmi \tilde z$ with
$\delta \equiv (\mu_g - \mu_f)/2$.
Then the zero-point condition is recast as 
\begin{align}
D(z_\text{zero};\mu_f) + D(z_\text{zero};\mu_g)= 2 \cosh L\delta \, \cosh \rmi L \tilde z_\text{zero}  +2\cosh L \hat m =0.
\end{align}
Denoting the ratio $r=\cosh L\hat m / \cosh L \delta$,
we easily find the zeros as
\begin{align}
L z_\text{zero} 
&=  \rmi L  \frac{\mu_f+\mu_g}{2} + L \tilde z_\text{zero}  
\notag \\
&=  \rmi L  \frac{\mu_f+\mu_g}{2} +   (2\ell +1) \pi +
\left \{ \begin{array}{l}
 \pm \cos ^{-1} r          \qquad  (\text{for  } r \leq 1) ,\\
 \pm \rmi \cosh^{-1}r  \qquad  (\text{for  } r >1) .\\
\end{array}
\right .
\end{align}
For $r \leq 1$ the $z_{\text{zero}}$'s  line up in the $z$ plane along a straight line parallel to the real axis
offset by $(\mu_f + \mu_g)/2$ to the imaginary direction.
For $r>1$ they appear with an equal separation $2\pi/L$ 
along the two straight lines separated by $2 \cosh^{-1} r$.
In both cases the modified model $Z_{f+g}$ has $2L$ zeros in total, just as many as the original model.
Setting $\mu_g$ large enough, we can make these zeros
(and the associated critical points, too)  appear with large imaginary axis in the complex $z$ plane.
We note, however, that one critical point on the imaginary part remains sitting at $\sinh z_{\text{c}} \sim {\rmi} / {\beta}$
(and another critical point on the $x=\pi$ axis)
for large $\mu_g$ with fixed $\mu_f$.

We show the thimble structure of $Z_{f+g}$ for $\mu_f=1$ and 2 with $\mu_g=3$ in Fig.~\ref{fig:DT-Thimble}, for example.
One recognizes that 
the original integration contour on the real axis $[-\pi,\pi]$ can be replaced with a set of thimbles which includes
the thimble starting at the critical point on the imaginary axis.
This thimble gives the most important contribution to the integral.
By increasing the $\mu_g$ value,
we can move the critical points associated to the zeros further upward 
so that the single thimble becomes equivalent to the original integration contour.


\end{document}